# Solving the Reswitching Paradox in the Sraffian Theory of Capital

*By* CARLO MILANA[+]

Birkbeck College, University of London

Correspondence: Carlo Milana, Department of Management, Birkbeck College, Malet Street, Bloomsbury, London WC1E 7HX. Email: c.milana@bbk.ac.uk

Fourth revision, 27th November 2019

Published in *Applied Economics and Finance*, Vol. 6, No. 6, November 2019

The possibility of the reswitching of techniques in Piero Sraffa's intersectoral model, namely the recurring capital-intensive techniques with monotonic changes in the interest rate, is traditionally considered as a paradoxical violation of the assumed convexity of technology putting at stake the viability of the neoclassical theory of production. The purpose of this paper is to demonstrate that this phenomenon is rationalizable within the neoclassical paradigm. Sectoral interdependencies can give rise to the well-known Wicksell effect, which is the revaluation of capital goods due to changes in relative prices. The reswitching of techniques is, therefore, the result of cost-minimizing technical choices facing returning ranks of relative input prices in full consistency with the pure marginalist theory of factor rewards. The proposed theoretical analysis is applied empirically to revisit various numerical examples presented in the literature.

## 1. Introduction

The use of the concept of marginal productivity in Piketty's (2014) *Capital in the Twenty-First Century* has provoked a critical debate (including Galbraith, 2014 and Solow, 2014) drawing on the so-called Cambridge controversy of the 1960s and Piero Sraffa's (1960) canonical book. The debate put at stake the traditional inverse relationship between the rate of interest and the capital

[+] The author is grateful to Alberto Heimler for his advice and continuous interest in discussing this work and Robert L. Vienneau for his critical remarks on an earlier version of this paper. The usual caveats apply.

intensity of production, particularly with the noted phenomenon of the recurrence of techniques of production over different ranges of the interest rate. The reappearance of the same methods of production over monotonic changes in the interest rate would suggest a paradoxical internal inconsistency of the marginalist theory of factor rewards. Because of such phenomenon, the theoretical factor reward based on marginal factor productivity has been put definitely in doubt after Samuelson's (1966, p. 578) recognition from the neoclassical camp that "it is quite possible to encounter switch points […] in which lower profit rates are associated with lower steady-state capital/output ratios'. By interrupting the monotonicity of the inverse relationship between the rate of interest and capital intensity, reswitching from "perverse' to well-behaved input demand in subsequent lower interest rates brought about perplexity about the traditional paradigm (Scazzieri, 2008a, 2008b). Samuelson (1966, p. 577) admitted his surprise in these terms:

> The reversal of direction of the (*i*, *NNP*) relation was, I must confess, the single most surprising revelation from the reswitching discussion. I had thought this relation could not change its curvature if the underlying technology was convex.

Since then, it has become customary to represent the "reswitching" phenomenon graphically by plotting, for each technique, the wage-interest relation (more precisely the trade-off relation between the real labor wage (say $w_L$) relative to the output price ($p$) and the interest rate ($r$)), which can be derived from the respective accounting equation of the cost of production. In **Figure 1**, the wage-interest curves of two alternative techniques $\alpha$ and $\beta$ define the level of the real wage for a given interest rate. By increasing the interest rate progressively, the system switches from $\alpha$ to $\beta$, at the interest rate $r_1$, and then switches back to $\alpha$, at the higher interest rate $r_2$.

The debate on such a phenomenon has been recently revived, especially regarding its economic significance and its relevance for aggregation in current macroeconomic analyses. This result is seen to have implications for the micro-foundations of macroeconomic models (Baqaee and Farhi, 2018), where the Sraffian criticism pointing to the noted inconsistency with the marginalist theory is still considered valid. As it will be demonstrated below in this paper, the Sraffian criticism mistakes the interest rate for the capital input price. This misconception leads to a wrong conclusion pointing to apparent internal inconsistency of the marginalist theory. It fails to note that, in such a paradox, a monotonic succession of the interest rate levels corresponds to a succession in opposite directions of the real capital-input price. As it will be shown below, in the case of **Figure 1**, the technical reswitching $\alpha$-$\beta$-$\alpha$ over the range of interest rate would correspond to a single technical switch $\alpha$-$\beta$ over the range of the *real rental prices* of capital goods. In fact, each technique brings about a linear relation between these real prices and real wage. The main proposition of the present paper is therefore that, *consistently with the marginalist theory of production and distribution, the reswitching of techniques can never occur in the coordinate space of the true real factor prices with any configuration of the Sraffian model*.[1]

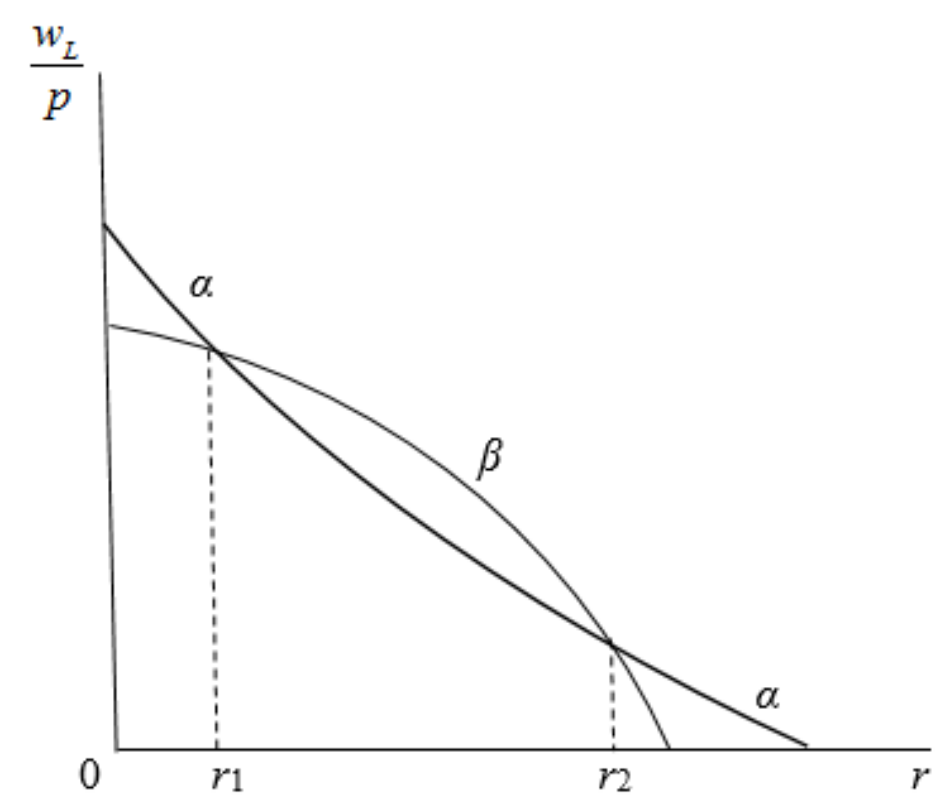

**FIGURE 1**. SRAFFIAN 'RESWITCHING OF TECHNIQUES' AND WAGE-INTEREST CURVES OF TWO ALTERNATIVE TECHNIQUES OF PRODUCTION $\alpha$ and $\beta$.

However, a consensus about the theoretical significance of reswitching has never been reached as demonstrated by continued discussions (Coen & Harcourt, 2003, Pasinetti, 2003, Garegnani, 2012, Schefold, 2013, Backhouse, 2014, Gram & Harcourt, 2017, Vienneau, 2017,

[1] Paul A. Samuelson, instead, initially suggested that each technique always brings about a linear relation between the *interest rate* and the real wage. Starting from this view, one of his PhD students, David Levhari, proposed a non-reswitching theorem (Levhary, 1965), which was later falsified in the *QJE* 1966 Symposium where Levhari and Samuelson's (1966) capitulation made the reswitching paradox accepted as a fact of life (see also Samuelson, 1966).

Lewin & Cachanosky, 2019). The reswitching and reverse capital deepening are seen as paradoxical phenomena, but they derive from technical choices under different financial conditions. The effects of the interest rate on relative prices are well known to financial engineers at least since Irving Fisher (1907, pp. 352-53; 1930, p.279) and Hayek (1931, 1941). Fisher, in his classic works on interest, was also aware of the possibility of reversing capital *value* in relation with interest (Samuelson, 1966, p. 581, fn. 2, Velupillai, 1975, 1995, Garrison, 2006), giving rise to the so-called Wicksell effects.

If a lower *interest rate* is faced by a "perverse" choice of a less (physical) capital-intensive technique, such a choice ceases to appear "perverse" if it is rationalizable with the consequent higher level of the *rental prices of capital goods*. It is straightforward to note that the reswitching of techniques disappears in the coordinate space of real factor prices by sorting the ratio of unit costs of production obtained with two alternative techniques. Hicks (1973, p. 45) was aware that reswitching is due to changes in the steady-state relative prices at different ranges of interest rate:

> A switch from one technique to another may be motivated by saving that is made in one of these directions; but in order to secure that saving, it may well be necessary—technically necessary—in some directions to incur some extra expense. The switch will nevertheless be profitable if the saving outweighs the expense. But the balance between this saving and this expense itself depends upon *prices* (represented, in the present model, by [the real wage] *w/p*); *if prices were different it might go the other way*. That is what produces the "re-switching' possibility. (Emphasis added and notation adjusted.)

Sraffian switch points in the real wage-interest plane occur when *all* relative prices (including the interest rate and real wage) are equalized between the two compared techniques. Although it is generally rare and small-sized, reswitching is still wrongly considered as the most important internal contradiction of the neoclassical marginal productivity theory. This internal contradiction is only apparent as it turns out to originate from a flawed interpretation. As this paper clarifies, this is another case of what Leontief's (1937) called "implicit theorizing" of Cambridge (UK) economists. It appears only when the analysis is carried out from the financial perspective. When the physical capital goods are instead taken into consideration within *explicitly* defined cost accounts, the reswitching of techniques invariably disappears. The dichotomy between financial and productive perspectives is well known: see, for example, Lerner (1953) and the seminal works of Morishima (1964, 1969) and Hicks (1965), who employed examples of both neoclassical and Sraffian models to deal with capital theory and economic growth (see also Gram, 1976, although with a Sraffian partisan view).

This paper aims at demonstrating that these two types of models complement rather than contradict each other by offering a fresh view of the real significance of the Sraffian reswitching based on a new general theorem. In light of these results, the Sraffian counterexamples of reswitching proposed in the literature are revisited by considering the various cases proposed in the literature. They include models where all the same commodities are produced with all techniques, and all commodities are basic in all techniques as well as cases where different commodities are specific to different techniques. Central to the analysis is Fisher's (1930, p. 131) and Wicksell's (1934) notion that interest rate change is part and parcel of changes in the structure of price ratios. As the debate for alternative investment projects had shown (see for example Alchian, 1955, Pitchford *et al*. 1958, Ramsay, 1970 among others), the discovery of *multiple* solutions for *internal rate of return* (or Keynes' *marginal efficiency of capital*) has become commonplace in their association with the *same* flow of net financial returns from a capital good. The latter coincides, in equilibrium, with the flow of rentals reflecting the marginal productivity of the capital asset. It will be shown that the erroneous interpretation of the reswitching paradox stemmed just from overlooking a similar correspondence between multiple combinations of interest rate and real wage–at different Sraffian switch points–corresponding to *relative factor prices* that are consistent with the marginal productivity theory. The same type of analysis proposed here can be easily extended to the Austrian models of intertemporal production costs by Pasinetti (1966) and Samuelson (1966).[2]

[2] The reswitching of techniques has been originally proposed as a paradoxical phenomenon without any direct reference to capital aggregation. Therefore, it is treated here without fully discussing its implication on aggregation theory within macroeconomic general-equilibrium models. References to the relation between the reswitching and

## 2. The reswitching paradox in the Sraffian model

One of the significant problems with Sraffa's model of prices is that, as Afriat (1987, p. 189) noted, "there is an obstacle to the application of the theory, *since the arithmetic of it is impossible"* (emphasis added) as there are too many conditions imposed by Sraffa on his system of prices. Afriat's criticism did not come alone. Samuelson (1962, 1966, 1975, 1983, 2000), Hicks (1965), Morishima (1966), Solow (1969, 1975), Stiglitz (1973, 1974), Sen (1974), Bliss (1975), Hahn (1975, 1982), and Mandler (1999a, 1999b, 2008) among many others, have shown perplexity about other aspects. Samuelson (2000, p. 113), in particular, referred to Sraffa's (1960) book as "a work in mathematical economics by an amateur, an autodidact. It has the properties of such. The book has more in it that the author knows. It is not the better for its imperfections."

Surprisingly, however, the fallacy of the interpretation of the reswitching of techniques as a violation of the neoclassical theory of production has rested unchallenged thus far. Dobb (1973, p. 252), considered Sraffa's (1960) demonstration of the possibility of […] "the double-switch of techniques" as his "most important contribution to a 'Critique of Economic Theory'." Indeed, Samuelson (2000, p. 117) himself praised the last seven pages dedicated to the choice of techniques in Sraffa (1960, Chap. XII) to "constitute the novelty of the work's contribution." He had previously claimed that, though this 100-page book "presents results that are compatible with marginalist theory or certain modern generalizations of that theory of the linear programming type, we have no right to indict Sraffa for being a marginalist" (Samuelson, 1961, p. 423). The present paper argues, instead, that the reswitching of techniques is not contradictory but entirely consistent with the neoclassical paradigm.

aggregation problems are the discussions started with Champernowne's (1953-1954) comment on Robinson's (1953-1954) famous article *The Production Function and the Theory of Capital,* further revisited by Brown (1980) and, more recently, Baqaee and Farhi (2018). The propositions on aggregation in this literature critically depend on the reswitching phenomenon.

In the opening of the chapter dedicated to the concept of paradox, *The Elements of Eloquence* by Forsyth (2013: Chap. 23) states:

> Paradoxes are remarkably hard to define, but you know one when you see one. Mathematicians, logicians, psychologists, sociologists and poets all compete for the word. They all think they own it. But this is untrue. For paradoxes are quite paradoxical."

Nonetheless, definitions of paradoxes have been attempted. As noted by Al-Khalili (2013), real paradoxes in science are statements that lead to circular and self-contradictory arguments or describe logically impossible situations. They are seen to be generally due to false assumptions or erroneous linking valid assumptions with wrong conclusions or, if starting from valid assumptions and using a correct logic, the right conclusions would appear contrary to the common sense arising from the narrow interpreter's vision (Sorensen, 2005; Sainsbury, 2009). Among the most famous paradoxes in economics, the Giffen's paradox suggests that certain goods have an upward sloping demand curve with nominal wealth held constant, thus violating the law of demand, a phenomenon that has been rationalized by extending the demand theory to the concept of inferior goods (Marshall, 1920: 109-110; Nachbar, 2008). Another case is the so-called "Leontief (1953) paradox" originated from the apparent contradiction of the Hecksher-Ohlin theorem in the pure theory of international trade on the basis of certain observed data (e.g. Woodland, 1982: 378-381; Gandolfo, 1987: 94-98). Still debated today, this paradox appears to be solvable as a discrepancy between pure theorems viewed as "controlled" (or restricted) thought experiments and "uncontrolled" (or unrestricted) real life of economic systems. The "reswitching" paradox was, instead, presented as an internal logical contradiction of the marginalist theory—for example, in the critical papers of the Symposium 'Paradoxes in Capital Theory' hosted in the special issue of November 1966 of *The Quarterly Journal of Economics*. As it will be shown in the present paper, the "reswitching paradox" turns out to be another case of deductive incoherence in flawed interpretations of the type called "implicit theorizing" found by Leontief (1937: 339) in the "the logical pattern used by [U.K.] Cambridge economists."[3]

Much of the confusion and debate have come from the definition of the capital input price. As already mentioned, in different occasions, Hicks (1965, p. 140, fn.1; 1979) recalled that it is the net earnings of the proprietors of an asset rather than the rate of interest that represents the price of capital service in the general case (see, for example, Petri, 2016 among recent discussions). It is well known, at least since Walras (1896), that the unit cost of using durable capital goods in production is represented by the *rental price of capital service* of which the rate of interest is a component along with the acquisition prices of the same capital goods. This was in line with the various studies of Wicksell (1893, 1901, 1934), who noted a discrepancy between the marginal productivity of capital goods and the rate of interest in the equilibrium of a social system. In a stylized model of the pure neoclassical theory of production, it turns out that, in the stationary equilibrium, the marginal productivity of capital goods is equal to the real price of capital services, not merely to the rate of interest. In other words, the marginal productivity of a capital good is found to be equal to the interest rate multiplied by a factor that may be larger or lower unity depending on the acquisition price of the capital good relative to the output price.[4]

In the Sraffian intersectoral model of production, the rate of interest may indeed interact in various directions with all relative prices. The intersectoral interaction between the rate of interest and relative input prices is generally highly non-linear. Different techniques yield equilibrium price solutions that may correspond to different levels of the interest rate. As it will be shown in the present paper, the price-taking producers may "re-switch" techniques as a cost-minimizing response to changes in relative input prices. [5]

[3] On the coherence theory of truth, see also Priest (2000, 2006a, 2006b).

[4] Pasinetti (1978, p.183) offered a schematic description of Wicksell's analysis. A more general treatment of the relation between marginal productivity of capital and interest rate, is contained in Leontief (1934), Metzler's (1950, 1951) and further discussed in Lerner (1953), who clarified the relations between the rate of interest and marginal private and social products of capital through the use of physical capital goods.

[5] Gallaway and Shukla (1976) recalled that the most profitable technique is not necessarily the one with the highest rate of profit for a given real wage (see Laibman and Nell, 1977, pp. 883-84 for a discussion on this point). Salvadori (1985) reached the same conclusion in the joint production case, but he claimed that this is due to the existence of

To be sure, initially, Joan Robinson (1953-1954) was very cautious when discussing the "curious possibility" of the reswitching of techniques pointed out to her by Ruth Cohen. In this view, the "perverse" behavior of the curve defined in the real wage-interest coordinate space, when it occurs, can be met only rarely and over a limited range. Reswitching was recognized by Champernowne (1953-1954) in his comment on her article and reaffirmed by Robinson (1956) and became more fully explicit in Sraffa's (1960) last seven pages. Many years later, Joan Robinson intervened again on the subject. She confirmed her views about the unimportance of reswitching relative to other major issues such as those regarding the existence of an aggregate pseudo production function (Robinson and Naqvi, 1967 and Robinson, 1969b, 1975a, 1975b). As she described it, *"[w]hat "reswitching" showed was that a higher real-wage rate is not necessarily associated with higher net output per head, and a lower rate of profit with a higher value of capital per man employed"* (Robinson, 1975b, p. 553; emphasis added). She noted that a good deal of exploration was needed before saying whether this phenomenon is a mere theoretical "rigmarole," or whether there is likely to be anything in reality corresponding to it. However, while the aggregation theory has reached a mature state in the mainstream neoclassical field, the remote possibility of a perverse behavior of the relation between capital intensity and the interest rate has remained the only major powerhouse argument of the Sraffian critics (Mas-Colell, 1989).

An inspiring *incipit* of the article by Robinson (1975a) is worth quoting:

> The story of what is known as the debate over the reswitching of the techniques is a sad example of how controversies arise between contestants who confront the conclusions of their arguments without first examining their respective assumptions. How is it possible to have a controversy over a purely logical point? When various theorists each set out their assumptions clearly, after eliminating errors, they can agree about what conclusions follow from what assumptions (*ibid*., p. 32).

joint production. However, none of these authors went as far as to overhaul the interpretation of the "reswitching" of techniques in terms of violation of the marginalist price-quantity behaviour.

.

## 3. Background concepts: "marginalist" values of capital goods

The Sraffian interpretation of reswitching originated in a historical context where the lack of reliable disaggregated data during the first half of the twentieth century has contributed to the adoption of aggregate economic models with the notable exceptions of Leontief (1941, 1953) and Afriat (1972, 2014) (on the latter, see also Afriat and Milana, 2009). The use of the aggregate production functions made the economists familiar with the concept of economy-wide optimal factor demands. In the particular case of a Clark-Ramsey economy producing one single commodity "by means of" the same commodity and labor in an aggregate economic model (where the output is partly re-usable as an input of production), it was customary to consider two aggregate inputs. Denoting the quantity and producer price levels of gross output of the economy respectively with $y$ and $p_y$, under the competitive equilibrium condition and constant returns to scale where supernormal profits are zero, equals the average total cost of production. Then, the optimality first-order conditions yield the marginal factor productivities to equal the real factor price in terms of the output price (see, for example, Intriligator, 1971, p. 191):

$$MPL \equiv \frac{\partial f(\mathbf{x})}{\partial x_L} = \frac{w_L}{p_y} \tag{1}$$

$$MPK \equiv \frac{\partial f(\mathbf{x})}{\partial x_K} = \frac{p_y(\delta + r)}{p_y} = \delta + r \tag{2}$$

where $x_L$ and $x_K$ are respectively quantities of labor and capital inputs, $\delta$ is the depreciation rate taking values within the interval $0 \le \delta \le 1$, and $r$ is the interest rate, whose equilibrium level is equal to the profit rate and determined in the financial market, $w_L$ is the labor wage rate assumed here to be paid *post factum*. The Appendix provides the derivation of the rental capital price $w_K = p_y(\delta + r)$ from the intertemporal cost minimization.

The textbook definition of the marginal rate of input substitution (*MRS*) subject to a neoclassical production function $f(\mathbf{x})$ (*ibid*., p. 193) is

$$MRS \equiv \frac{dx_L}{dx_K}\bigg|_{isoquant} = -\frac{MPK}{MPL} \tag{3}$$

Therefore, given (1) and (2), in the model with a homogeneous output,

$$-\frac{dx_L}{dx_K}\bigg|_{isoquant} = \frac{w_K}{w_L} \tag{4}$$

The equality $\frac{w_K}{w_L} = \frac{(\delta + r)}{w_L / p_y}$ has led to the ratio of $(\delta + r)$ to $w_L / p_y$, as an indicator of the price of capital input relative to the real labor wage. Such a customary formulation became mistaken when it was extended to the interindustry models, and the error has leaked into many studies bringing about paradoxical meanings of the results.

In an interindustry model with heterogeneous outputs, the definition of the marginal rate of substitution leads to the equilibrium equality

$$-\frac{dx_L}{dx_K}\bigg|_{isoquant} = \frac{w_K}{w_L} = \frac{p_K(\delta + r)}{w_L} \tag{5}$$

The correct expressions for the real wages of capital and labor inputs are in fact $\frac{p_K(\delta + r)}{p_y}$ and $w_L / p_y$ respectively. The real rental of the self-produced capital good is instead equal to $\delta + r$ as, in this case, $p_K = p_y$. An equivalent discrete form of the (negative) marginal rate of input substitution given by (5) applies to Sraffa's (1960, Chap. XII) model of a cost-minimizing choice over a finite set of alternative techniques of production. However, in an intersectoral model and even in macroeconomic models, where the output is used for both immediate consumption and accumulation purposes, the relative acquisition prices are affected by changes in income distribution. The output price is, therefore, an aggregate of the respective price indexes in consumption and investment activities, that is $p_y = g(p_C, p_K)$ with $p_y \neq p_C \neq p_K$. The solution of the model takes account of the interrelation between the interest rate and the relative prices (as, for example, in Harcourt, 1970, p. 45). In the next section, this price interrelation also applies to the intersectoral models where monotonic changes in the interest rate affect relative prices in a *non-linear* way while playing a pivotal role in equilibrium solutions.

## 4. Accounting for prices in the Leontief-Sraffa model

The simplest version of the Sraffian model of production of commodities "by means" of commodities is that of two sectors using two or three inputs originally proposed by Samuelson (1962, pp. 204-205). In most examples of the two-sector two-input model, the focused sector produces the consumption good using labor while the capital goods are acquired *ex-ante* (at the start of the current period), from the second sector producing the capital good using labor and a quantity of its own output, acquired *ex-ante* from itself.

The Sraffian solutions are clearly seen from the accounting system of a generalized Leontief-Sraffa type model. Let us consider a simplified production system with the following characteristics. All commodities can be produced over one certain period at constant returns to scale with no joint production, out of themselves, and out of one or more primary factors produced in the preceding periods. In a fully competitive equilibrium of the current period, there are no supernormal profits on production activity where the rate of interest is the same in all sectors. In such conditions, the output price equals the average total cost of production. The general system of price accounting equations can be expressed in matrix form as

$$\mathbf{p} = w_L \mathbf{a}_0 + \mathbf{pB}(\hat{\boldsymbol{\delta}} + \hat{\mathbf{r}}) + \mathbf{pA} \qquad (6)$$

where $\mathbf{p}$ is the *n*-order row vector of output prices; is the *ante factum* labour wage rate; is the labour wage rate paid *post factum,* which is equivalent to the present value of the labor wage paid *ante factum* $w_0$, so that $w_L = w_0(1+r)$ (differently from most numerical examples, the original

Sraffa's, 1960 model does not consider the labor wage paid *ante factum*); $\mathbf{a}_0$ is the *n*-order row vector of direct input-output coefficients of labor*;* **I** is the $(n \times n)$-order unit matrix**;** **r** is the *n*-order vector of internal rates of return or rates of profit; $\boldsymbol{\delta}$ is the *n*-order vector of non-negative depreciation rates of capital goods (the hat ^ indicates the transformation of a vector in a diagonal matrix); **A** is the $(n \times n)$-order Leontief matrix of direct input-output coefficients for intermediate circulating goods produced and consumed during the current period of production; **B** is the $(n \times n)$-order Leontief matrix of input-output coefficients for the services of capital goods pre-existing at the start of the current period of production.

In Sraffa's model of production of commodities by means of commodities where all $\delta_i = 1$, the accounting equation (6) of an economy in equilibrium condition with all sectors scoring the same profit rate *r* can be solved in one single step with the following reduced form if the Hawkins and Simon condition on the viability of the system is satisfied:

$$\mathbf{p} = w_L \mathbf{a}_0 [\mathbf{I} - \mathbf{A} - (1+r)\mathbf{B}]^{-1} \tag{7}$$

However, Sraffa's analysis can be referred to a two-step solution of the system of price accounting equations (6). The first step computes the price components defined at the level of the Leontievian "vertically integrated sectors" containing all the interindustry transactions occurring in the current period to produce the final quantities of commodities. This solution is obtained by taking account of the interactions between the sectoral prices during the same current period and considering the prices of pre-existing factor inputs as predetermined variables:

$$\begin{aligned} \mathbf{p} &= w_L \mathbf{a}_0 (\mathbf{I} - \mathbf{A})^{-1} + \mathbf{p}(1+r)\mathbf{B}(\mathbf{I} - \mathbf{A})^{-1} \\ &= w_L \mathbf{a}_L + \mathbf{p}(1+r)\mathbf{A}_K \\ &= \mathbf{w}\mathbf{A}_T \end{aligned} \tag{8}$$

where $\mathbf{a}_L \equiv \mathbf{a}_0(\mathbf{I}-\mathbf{A})^{-1}$ is the $n$-order row vector of Leontief's *direct and indirect* input-output requirements of labor inputs; $\mathbf{A}_K \equiv \mathbf{B}(\mathbf{I}-\mathbf{A})^{-1}$ is the $(n \times n)$-order matrix of Leontief's *direct and indirect* input-output coefficients for inputs of capital goods services; $\mathbf{w} \equiv [w_L\ \mathbf{w}_K]$ is the $(n+1)$-order row vector of wage rates for inputs of labor and capital good services; $\mathbf{w}_K \equiv \mathbf{p}(1+r)$ is the $n$-order row vector of rental prices or user cost of capital goods; $\mathbf{A}_T \equiv \begin{bmatrix} \mathbf{a}_L \\ \mathbf{A}_K \end{bmatrix}$ is the $[(n+1)\times n]$-order matrix of Leontief's direct and indirect input-output requirements for total labor and capital goods.

In the second step, the reduced form of the Sraffa's price model represented by the second line of (8):

$$\mathbf{p} = w_L \mathbf{a}_L [\mathbf{I} - (1+r)\mathbf{A}_K]^{-1} \tag{9}$$

The prices **p** are positive if $0 \leq r \leq R$, where $R$ is the maximum profit rate attainable in the production system, that is $R = (1/\lambda) - 1$, and $\lambda$ represents the dominant eigenvalue of $\mathbf{A}_K$ (see, for example, Pasinetti, 1977, pp. 95-97).

The price accounting system is made of $n$ equations with the $(n + 2)$ price-type variables $\mathbf{p}, w_L, r$. Let all prices and the wage rate be divided through (9) by one arbitrary output price, say $p_j$. Moreover, let $w_L / p_j$, or $r$, or a normalized output price, say $p_i / p_j$, be pre-determined, then the Sraffian system of price equations can be respectively solved to find the $n$-tuple of either $(\frac{1}{p_j}\mathbf{p},\ r)$, or $R = (1/\lambda) - 1$, or the rest of $(n-2)$ normalized output prices along with $(\frac{w_L}{p_j}, r)$. In the Sraffian approach, the output prices and labor wage are usually expressed in real terms as ratios to the output price of one focused commodity, while the level of the real wage rate or rate of profits is conjecturally fixed.[6]

Hence, the technology can autonomously determine all relative prices except one, which can be chosen arbitrarily. It is noteworthy that the second-step reduced form (9) of (6) gives rise to the accounting expression of the price decomposition of the so-called Sraffian "sub-systems", which are conceptually different from the price decomposition of the Leontievian "vertically integrated sectors" corresponding to the first-step reduced form (8) of (6). Some authors, for example, Gram (1976), Pertz and Teplitz (1979), considered (8) and (9) as two different alternative views of price determination. In the following discussion of the reswitching paradox, *both* (8) and (9) forms provide, instead, complementary information that is needed to satisfy Sraffa's (1960, Chap. XII) requirements for identifying the switch points between alternative techniques of production. In these switch points, the techniques coexist with all the *n*-tuples of relative prices plus the real wage and the interest rate being equal. Contrary to the usual discussions in the reswitching debate, which have been generally centered on the reduced form (9), the sufficient conditions for the existence of genuine reswitching points need to be checked also on the full range of all relative prices by exploring more directly the structural form (8). In the case where each technique is part of a book of "blueprints" by producing and using specific types of capital goods, the matrices $\mathbf{a}_L$ and $\mathbf{A}_K$ can be augmented in the way described by Sraffa (1960, pp. 82-83) and Pasinetti (1977, pp. 165-167) to enable comparisons. However, as it will be noted, this algebraic arrangement leaves the numerical solutions unchanged.

[6] Sraffa defined also a theoretical "standard commodity" to be used as a *numéraire* whose computed price indeed depends on income distribution, but is invariant with respect to changes in the relative prices of other commodities. The price of the standard commodity corresponds to a weighted average of commodity prices where the weights are the elements of the eigenvector of the matrix $\mathbf{A}_K$. However, the dependence of movements of this particular price on the profit and real wage rates, support the abovementioned Afriat's criticism about the arithmetic impossibility of the simultaneous determination of a price "ultimately equal to the labour that has gone into making it" and being also the result of a particular income distribution. More recent discussions on the meaning of Sraffa's "standard commodity", on which there is not yet a commonly accepted view, include those of Bellino (2004), Baldone (2006), and Wright (2014, 2017).

The $j$th sub-system features the following real labor wage as a function of the profit rate for a given technology:

$$\frac{w_L}{p_j} = \frac{1}{\mathbf{a}_L[\mathbf{I} - (1+r)\mathbf{A}_K]^{-1}\mathbf{e}_j} \tag{10}$$

where $\mathbf{e}_j$ is a column vector with all its elements equal to zero except the $j$th one, which is equal to unity. The capital input prices expressed in terms of the $j$th commodity are obtained as

$$\mathbf{w}_K \frac{1}{p_j} = \mathbf{p}\frac{(1+r)}{p_j} = \frac{w_L}{p_j}\mathbf{a}_L[\frac{1}{(1+r)}\mathbf{I} - \mathbf{A}_K]^{-1} \tag{11}$$

Dividing (11) through by $\frac{w_L}{p_j}$ yields

$$\frac{1}{w_L}\mathbf{w}_K = \mathbf{a}_L[\frac{1}{1+r}\mathbf{I} - \mathbf{A}_K]^{-1} \tag{12}$$

which, given (8), (11), and (12), is equivalent to

$$\frac{1}{w_L}\mathbf{w}_K = \frac{1+r}{w_L}\mathbf{w}\mathbf{A}_T \tag{13}$$

## 5. The Sraffian price system from the perspective of linear programming

In order to clarify further the meaning of the rental prices of capital goods used in the system (8) and their relationship with the interest rate, the third line of (8) can be complemented with the objective function specifying the assumption of cost minimization throughout the economy in the following classical programming problem

$$C(\mathbf{w}) = Min_{\mathbf{w}}\ \mathbf{w}\cdot\mathbf{v} \tag{14}$$

sub to

$$\mathbf{p} = \mathbf{w}\mathbf{A}_T$$

where $\mathbf{v}$ is a given (column) vector of primary inputs.

The foregoing minimization problem has the dual counterpart of the quantity maximization

$$\begin{aligned} &R(\mathbf{f}) = Max_{\mathbf{f}}\ \mathbf{p}\cdot\mathbf{f} \\ &\text{sub to} \\ &\mathbf{A}_T\mathbf{f} = \mathbf{v} \end{aligned} \tag{15}$$

where $\mathbf{f}$ is a given (column) vector of final outputs.

The Lagrangian functions of problems (14) and (15) are, respectively

$$L(\mathbf{w},\ \boldsymbol{\lambda}_C) = C(\mathbf{w}) + (\,\mathbf{p} - \mathbf{w}\mathbf{A}_T\,)\,\boldsymbol{\lambda}_C \tag{16}$$

$$L(\boldsymbol{\lambda}_R, \mathbf{f}) = R(\mathbf{f}) + \boldsymbol{\lambda}_R\,(\,\mathbf{v} - \mathbf{A}_T\mathbf{f}\,)$$

The conditions for a stationary point of $L(\mathbf{w},\ \boldsymbol{\lambda}_C)$ are

$$\begin{aligned} &\nabla_{\mathbf{w}^*} L(\mathbf{w}^*, \boldsymbol{\lambda}_C^*) = \nabla_{\mathbf{w}^*} C(\mathbf{w}^*) - \mathbf{A}_T\boldsymbol{\lambda}_C^* = \mathbf{0} \\ &\nabla_{\boldsymbol{\lambda}_C^*} L(\mathbf{w}^*, \boldsymbol{\lambda}_C^*) = \mathbf{p} - \mathbf{w}^*\mathbf{A}_T = \mathbf{0} \end{aligned} \tag{17}$$

and those of $L(\boldsymbol{\lambda}_R, \mathbf{f})$ are

$$\begin{aligned} &\nabla_{\mathbf{f}^*} L(\boldsymbol{\lambda}_R^*, \mathbf{f}^*) = \nabla_{\mathbf{f}^*} R(\mathbf{f}^*) - \boldsymbol{\lambda}_R^*\mathbf{A}_T = \mathbf{0} \\ &\nabla_{\boldsymbol{\lambda}_R^*} L(\boldsymbol{\lambda}_R^*, \mathbf{f}^*) = \mathbf{v} - \mathbf{A}_T\mathbf{f}^* = \mathbf{0} \end{aligned} \tag{18}$$

Solving the $2n + 1$ equations (17) yields the solutions for $2n + 1$ unknowns: $n + 1$ instruments (input prices) $\mathbf{w}^*$ and $n$ Lagrangian multipliers $\boldsymbol{\lambda}_C^*$. Similarly, simultaneously solving the $2n + 1$ equations (18) yields the solutions for other $2n + 1$ unknowns: $n + 1$ Lagrangian multipliers $\boldsymbol{\lambda}_R^*$ and $n$ instruments (final outputs) $\mathbf{f}^*$. Moreover, $\nabla_{\mathbf{w}^*} C(\mathbf{w}^*) = \mathbf{v}$ and $\boldsymbol{\lambda}_C^* = \mathbf{f}^*$. Similarly $\nabla_{\mathbf{f}^*} R(\mathbf{f}^*) = \mathbf{p}$ and $\boldsymbol{\lambda}_R^* = \mathbf{w}^*$. Therefore, under the assumptions made, the equality of optimal total cost expenditure and revenue is attained, that is $R^* = C^*$, which follows from $(\mathbf{p} - \mathbf{w}^* \mathbf{A}_T)\boldsymbol{\lambda}_C^* = \mathbf{0}$, implying $\mathbf{pf}^* = \mathbf{w}^* \mathbf{A}_T \mathbf{f}^*$, and $\boldsymbol{\lambda}_R(\mathbf{v} - \mathbf{A}_T \mathbf{f}^*) = \mathbf{0}$, implying $\mathbf{w}^* \mathbf{v} = \mathbf{pf}^*$.

If the resources **v** are allowed to vary, then using the resulting modified Lagrangian function derived from problem (15) yields the marginal optimal values of the inputs of production **v**. These values are obtained by differentiating the Langrangian with respect to :

$$\nabla_{\mathbf{v}} L(\mathbf{v}, \boldsymbol{\lambda}_R^*, \mathbf{f}^*) = \boldsymbol{\lambda}_R^* = \mathbf{w}^* \tag{19}$$

the demonstration of which, omitted here to save space, can be found, for example, in Intriligator (1971, pp. 36-38). In view of (8), the optimal input price vector $\mathbf{w}^* \equiv [w_L^* \ \mathbf{w}_K^*]$ with the rental prices $\mathbf{w}_K^* = \mathbf{p}_K^*(1+r)$ for the inputs of capital goods services is consistent with the definition of capital rental price dating back at least to Walras (see also the Appendix for an alternative demonstration)[7]. As they measure the sensitivity of the objective value to the marginal changes in the respective resource quantities, they are often called "shadow prices".[8] The sensitivity analysis can be extended to the changes in the input-output coefficients of the matrix $\mathbf{A}_T$ as in the discussion provided in the following section.

## 6. The Cost-Minimizing Choice Over Alternative Methods of Production

Regarding two different methods of production, Sraffa (1960, p. 98) claimed:

> Two different methods of producing the same basic commodity can only co-exist at the points of intersection (that is to say, at those rates of profits at which the prices of production by the two methods are equal), since the two economic systems (which are respectively characterized by the two methods, but are alike in every other respect) will at such points necessarily have also the same commodity-wage and the same system of relative prices.

[7] When labor is not binding as in the simplest steady-state model where full employment is not imposed, following Ricardo an exogenous non-zero subsistence wage is imposed from outside the system. This is the so-called "Fixwage assumption' considered in many Sraffa-type models with a perfectly elastic supply of labor, where the dominant technique, the rate of return from that technique are determined from the technology frontier (Hicks 1973, p. 49).

[8] Since Sraffa assumed cost-minimizing behavior, the explicit introduction of the objective of cost-minimization in his intersectoral price model brings about the neoclassical rental prices of capital goods (earlier statements are given, for example, by Bruno, 1969, p. 47 and Salvadori, 1982).

DEFINITION 1*: The Sraffian point(s) of intersections of two methods of production, say method I and method II, are defined as those satisfying the following requirements simultaneously:*

*(i) the interest (or profit) rate* $r = r^*$ *where the numerical value(s) of r* is (are) the solution(s) of the following equality of the real wage equation for the focused sector:*

$$\mathbf{a}_L^I[\mathbf{I}-(1+r)\mathbf{A}_K^I]^{-1}\mathbf{e}_j = \mathbf{a}_L^{II}[\mathbf{I}-(1+r)\mathbf{A}_K^{II}]^{-1}\mathbf{e}_j \quad (20)$$

*(ii) the numerical value(s) of* $r^*$ *satisfying (20) should also satisfy the consistency requirement that the two methods yield the same system of relative prices in a genuine switch point, that is*

$$\mathbf{a}_L^I[\mathbf{I}-(1+r^*)\mathbf{A}_K^{II}]^{-1}\mathbf{e}_i - \mathbf{a}_L^{II}[\mathbf{I}-(1+r^*)\mathbf{A}_K^{II}]^{-1}\mathbf{e}_i = 0; \quad \forall i \neq j \quad (21)$$

*which, using (8), is equivalent to*

$$\mathbf{w}^*(\mathbf{A}_T^{II} - \mathbf{A}_T^I)\mathbf{e}_i = 0; \quad \forall i \neq j \quad (22)$$

A problem may arise using only the real wage-profit rate curves in search of the identification of the points of intersection "where the prices of production by the two methods are equal" (Sraffa, 1960, p. 102, Fig. 8). In general, the intersection points in the 2-dimensional coordinate wage-profit space do not map into intersection points of the alternative techniques in the *n*-dimensional coordinate space of the real input prices. In other words, equations (20) and (21)-(22) in general do not hold simultaneously.

**Figure 2** represents the case of the two sector-two factor-two technique model (where $\delta$ = 0) showing two switch points of the respective wage-profit curves of the two methods, corresponding to different relative prices in the left-hand side quadrant hosting the respective linear real-factor-price curves. It is interesting to note that this configuration does not exhibit the Sraffian switching point defined as the locus where the two techniques yield the same relative input prices *and* the same profit for a given real wage. Moreover, the upper envelope of the two intersecting curves in the left-hand side quadrant is the true real-factor-price frontier. This frontier takes shape in the multi-dimensional coordinate space of real factor prices which, in the Sraffian literature, has been always replaced with the upper envelope of the real wage-interest curves leading the Cambridge controversy to mistaken results[9].

A different situation can now be considered in the real factor-price frontier of a production system with the two commodities, three inputs, and two techniques. One or multiple intersection points can occur between the two techniques in the 2-dimensional coordinate space of real wage and profit rate. These intersection points may not map into corresponding points of the locus of a straight line, which is shown in **Figure 3**, connecting the points A and B. Sraffa himself made clear that only if the number of alternative techniques is *lower* than the number of factor inputs, then the reswitching of techniques is possible. Only in such a case may all relative prices and the profit rate be equal in multiple points across the alternative techniques for a given labor wage. By contrast, if the number of alternative techniques is equal to or higher than the number of factor inputs, then the reswitching of techniques is not possible.

[9] Mistaken results derived from confusing the wage-interest upper frontier with the true real-factor price frontier in both the formulation of Levhari's (1965) non-reswitching theorem based on Samuelson's (1962) "surrogate production function" and its confutation by Garegnani (1966), Bruno *et al.* (1966), Levhari and Samuelson (1966), Morishima (1966). See also Pasinetti (1977, pp. 169-173).

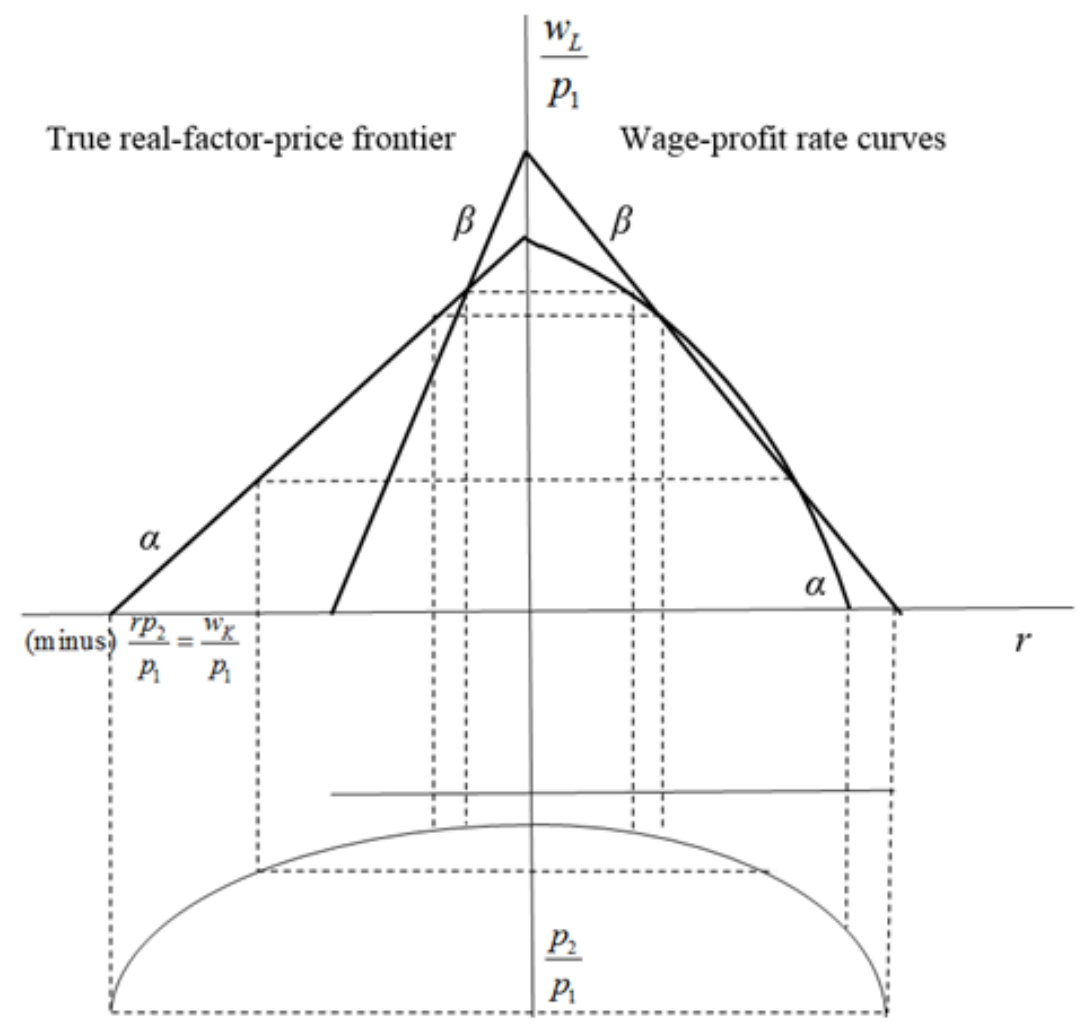

**FIGURE 2**—REAL FACTOR-PRICE FRONTIER AND REAL WAGE-PROFIT RATE CURVES IN A TWO SECTOR-TWO FACTOR-TWO TECHNIQUE MODEL WITH THREE INPUTS AND TWO TECHNIQUES.

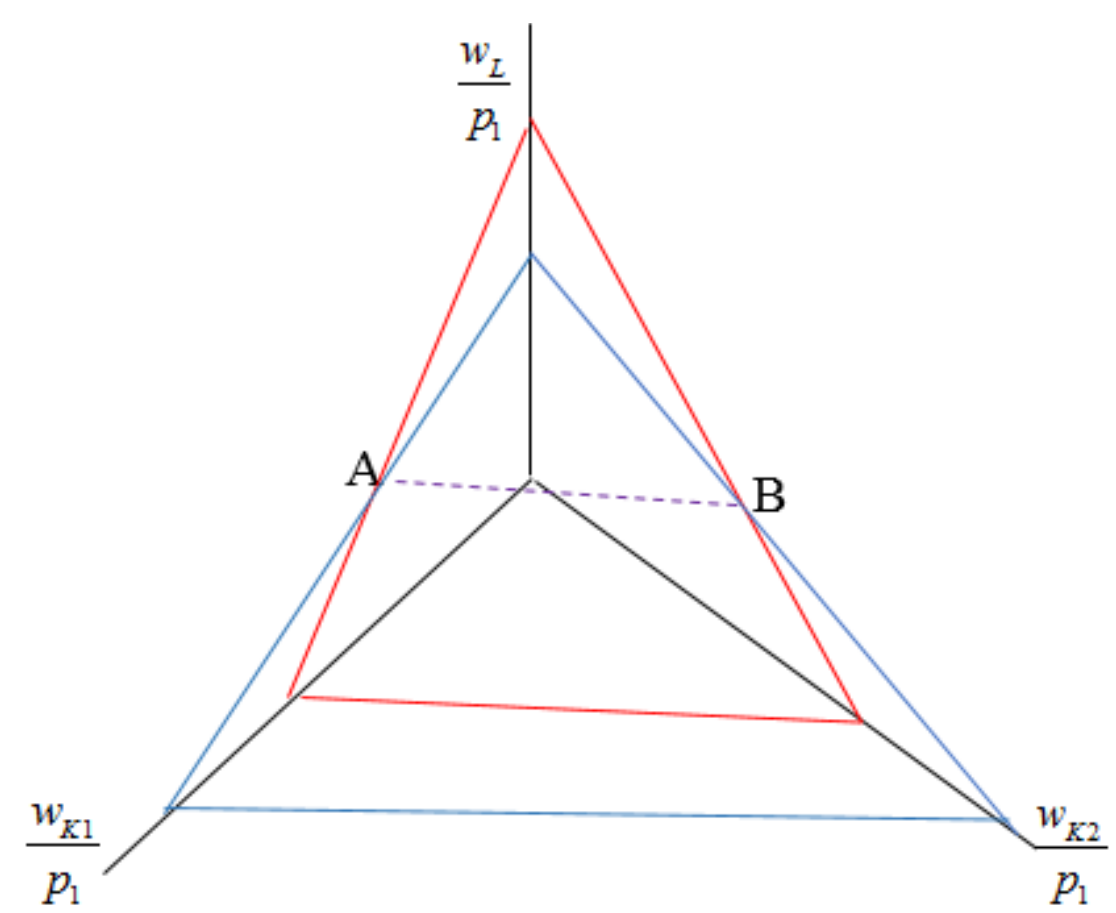

**FIGURE 3**—REAL FACTOR-PRICE FRONTIER IN A MODEL WITH THREE INPUTS AND TWO TECHNIQUES

He argued in the following terms:

> This co-existence [of alternative methods of production] is possible because with k basic equations (representing $k$ methods of production) and $k$+1 unknowns (representing $k$ – 1 [relative] prices, the wage $w_L$ and the rate of profits $r$ [with the total number of inputs being equal to $k + 1$]) there is room for one more basic equation (or method of production) even though it does not bring with it an additional product and an additional price. With $k + 1$ methods of production [with the number of methods being equal to the number of unknown real input prices], however, it is no longer possible to vary at will the rate of profits, its level is now fully determined. At any other level of the rate of profits the two methods are incompatible, and the two distinct systems to which they belong have no point of contact. (Sraffa, 1960, p. 90.)

The last Sraffa's foregoing passage can be described mathematically in the following cases.

### A. *The case of the number of techniques being equal to the number of factor inputs*

With the number of the alternative techniques being equal to the number of the factor inputs, the real factor prices are univocally determined provided the coefficient matrix is not singular. Let us consider the general case of $n$-sector Sraffian model with ($n + 1$) inputs (commodity inputs plus one labor input) and ($n + 1$) different techniques with each sector supplying its own product as input to the other sectors and to itself. Given the third line of (8), let the system of the real factor-price equations of the focused sector, say the first one, be represented as follows.

$$\begin{aligned}
&a_{L1}^{I}\frac{w_L}{p_1}+a_{11}^{I}\frac{w_{K1}}{p_1}+\ \dots\ +a_{n1}^{I}\frac{w_{Kn}}{p_1}=1\\
&a_{L1}^{II}\frac{w_L}{p_1}+a_{11}^{II}\frac{w_{K1}}{p_1}+\ \dots\ +a_{n1}^{II}\frac{w_{Kn}}{p_1}=1\\
&\dots\dots\dots\dots\dots\dots\dots\dots\dots\dots\dots\dots\\
&a_{L1}^{(n+1)}\frac{w_L}{p_1}+a_{11}^{(n+1)}\frac{w_{K1}}{p_1}+\ \dots\ +a_{n1}^{(n+1)}\frac{w_{Kn}}{p_1}=1
\end{aligned}\tag{23}$$

Recalling that $w_{K1}/p_1 = (1+r)$, the solution in matrix form is

$$\begin{bmatrix}\dfrac{w_L}{p_1} & (1+r) & \dots & \dfrac{w_{Kn}}{p_1}\end{bmatrix}=\begin{bmatrix}1 & 1 & \dots & 1\end{bmatrix}\begin{bmatrix}a_{L1}^{I} & a_{L1}^{II} & \dots & a_{L1}^{(n+1)}\\ a_{11}^{I} & a_{11}^{II} & \dots & a_{11}^{(n+1)}\\ . & . & . & .\\ a_{n1}^{I} & a_{n1}^{II} & \dots & a_{n1}^{(n+1)}\end{bmatrix}^{-1}\tag{24}$$

provided the matrix to be inverted is not singular. In this case, there is no degree of freedom for $r$ or $w_L / p_1$ with an income distribution fixed over different technical conditions. As Sraffa pointed out in the above-reported text, the solution is unique implying that reswitching is impossible.

In correspondence to the genuine switch point, the equation system (23) can be rearranged in order to relate the relative input prices to the sensitivity analysis of technical coefficients based on the stationary conditions (17). Starting with a binary comparison of techniques, say *I* and *II*, which at a switch point coexist with equal relative unit costs and prices, taking the difference of the respective cost equations and rearranging yield the following single equation in the *n* relative input prices for the focused sector:

$$(a_{L1}^{II} - a_{L1}^{I}) + \frac{1}{w_L^*}\mathbf{w}_K^*(\mathbf{a}_{K1}^{II} - \mathbf{a}_{K1}^{I}) = 0 \tag{25}$$

Applying the transitivity property of index numbers having common relative prices as weights, the whole system of binary comparison of *n* techniques at the same switch point yields

$$\begin{aligned} &\frac{1}{w_L^*}\mathbf{w}_K^*(\mathbf{a}_{K1}^{II} - \mathbf{a}_{K1}^{I}) = -(a_{L1}^{II} - a_{L1}^{I}) \\ &\frac{1}{w_L^*}\mathbf{w}_K^*(\mathbf{a}_{K1}^{III} - \mathbf{a}_{K1}^{II}) = -(a_{L1}^{III} - a_{L1}^{II}) \\ &\ldots\ldots\ldots\ldots\ldots\ldots\ldots\ldots\ldots\ldots \\ &\frac{1}{w_L^*}\mathbf{w}_K^*(\mathbf{a}_{K1}^{n^{th}} - \mathbf{a}_{K1}^{n^{th}}) = -(a_{L1}^{n^{th}} - a_{L1}^{n^{th}}) \end{aligned} \tag{26}$$

Solving for the rental rates of capital goods relative to wage in the common switch point yields

$$\frac{1}{w_L^*}\mathbf{w}_K^* = \mathbf{c}_1 \cdot (\mathbf{I} - \mathbf{D})^{-1} \tag{27}$$

where $\mathbf{c}_1$ is a (row) vector with elements $c_{i1} \equiv -\Delta a_{L1} / \Delta a_{Ki1} \quad \forall i$ , and $\mathbf{D}$ is a matrix with elements $d_{ss} = 0 \ \ \forall s$ and $d_{st} = -\dfrac{a^{\beta}_{Kt1} - a^{\alpha}_{Kt1}}{a^{\beta}_{Ks1} - a^{\alpha}_{Ks1}}$ where $s \neq t; \ s, t = 1, ..., n; \quad \beta \neq \alpha; \ \beta, \alpha = I, II, .., n^{th}$.

With only one capital good in the economy, the solution in terms of relative input prices is $\dfrac{w^*_K}{w^*_L} = -\dfrac{a^{II}_{L1} - a^{I}_{L1}}{a^{II}_{K1} - a^{I}_{K1}}$, which corresponds to (4) in the case of a continuous spectrum of techniques of production. The marginal productivity principle follows from the assumption of minimization of costs, non-joint production, and constant returns of scale in a competitive stationary equilibrium. The resulting zero profits result from the value of the output being equal to the total cost of production yielding the relative factor price equations (26) based on discrete marginal product ratios.[10] The Sraffian model thus yields the counterpart of the (negative) marginal rate of substitution (*MRS*) of capital and labor inputs defined, in the continuous case, by (3), (4), and (5) . Dorfman, Samuelson, and Solow (1958, p. 403) described the equivalence of relative prices with the *MRS* in the discrete models in these terms:

> No unique marginal rate of substitution exists—the slope of the efficiency frontier is different, depending on the direction we take. We can imagine these different slopes as putting limits to "the" *MRS*. We also know that at a corner or edge there will be more than one supporting plane, i.e., more than one associated set of *p*'s. (at a really flat place on the boundary, the supporting plane is unique and literally coincides with the boundary.) But *all* these supporting planes will have slopes within limits set for the *MRS*. So that even at the corners the generalized correspondence between associated *p* ratios and *MRS*'s persists. (Emphasis in the original.)

[10] As noted rightly by Bliss (1975, p. 94), "The importance of remembering that marginal concepts are not primary, but follow upon the basic postulates of maximization, is that it guides us when the specification of marginal equations is unclear back to the postulated maximization for an answer".

A real factor price at the corner "lies between the left- and the right-hand marginal product of the factor of production" (Bliss, 1975, p. 109).

At the intersection point of cost budget lines, the relative input prices are, in fact, the same with two techniques:

$$\begin{aligned} \mathbf{w}_{K}^{*} &= \mathbf{p}^{(I)*}(1+r^{*}) \\ &= \mathbf{p}^{(II)*}(1+r^{*}) \end{aligned} \qquad (28)$$

As Sraffa (1960, p. 90) himself noted, there is a unique vector of capital input prices and there is no reswitching in the general case where the number of alternative techniques is not lower than the number of commodity inputs and the rate of profit is uniquely determined.

The non-linear relationship between profit rate and relative input prices in terms of labor is at the heart of the apparent paradox of the reswitching of techniques, which may appear within intervals of possible values of the given profit rate or labor wage. The non-linear relationship between $r$ and relative prices implies that the profit rate and the capital input rentals in terms of labor may not change proportionally. The degree of non-linearity of such relation depends strictly on the rank of the Sraffian matrices being equal to the number of the producing sectors. In a two-sector model, the rental price relative to labor cost is a quadratic function of the profit rate. In general, with $n$ sectors, such a relationship has at most an $n$-degree polynomial form.[11]

With the two-sectors, two inputs, two-techniques, the solution (24) becomes

$$\begin{bmatrix} \frac{w_L}{p_1} & \frac{w_K}{p_1} \end{bmatrix} = \begin{bmatrix} 1 & 1 \end{bmatrix} \begin{bmatrix} a_{L1}^{I} & a_{L1}^{II} \\ a_{K1}^{I} & a_{K1}^{II} \end{bmatrix}^{-1} \qquad (29)$$

[11] For example, Schefold's (1976) analysis of he relative prices as a function of the rate of profit remained in the realm of the output prices within the Sraffian framework, whereas Gram (1976), offered a comparison of the two model solutions, but his analysis maintained the Sraffian interpretation of the real wage-profit rate relation as the real factor-price curve, but he did not clarify that this curve was the reduced form of the same model.

In stationary equilibrium, the equality $\frac{w_K}{p_1} = \frac{(1+r)p_2}{p_1}$ holds. There is no degree of freedom of *r* provided the relative price $p_2 / p_1$ has a level satisfying the technological condition (or vice versa). If this prototype model has common levels of $w_L / p_1$ and $w_K / p_1$, then this is generally achieved in correspondence with different levels of *r* using the alternative technique. This implies and is implied by the fact that the two techniques will generally differ in the relative price $p_2 / p_1$ in such a solution. Conversely, if the two techniques lead to two common levels of $w_L / p_1$ and *r,* then they generally differ in the relative rental price $w_{K1} / p_1$ and relative price $p_2 / p_1$. This excludes reswitching as defined by Sraffa (1960, p. 90). Common levels of $w_L / p_1$ and $w_K / p_1$ are achieved in correspondence with the same level of *r* only by a fluke. As Bruno *et al.* (1966) declared, reswitching is impossible in the presence of "only one capital good in the system." They wrote:

> Can we get reswitching if all activities use the same capital good? The answer to this question turns out to be negative, and we have the following theorem:
>
> THEOREM*: In a two-sector economy with many alternative independent techniques for producing the two goods, if there is* only one *capital good in the system, reswitching cannot occur.* (*Ibid*., p. 536, emphasis in the original.)

This contention, however, turns out to be misleading as the impossibility of reswitching also arises in the general case with more than one capital good. Sraffa (1960, p. 90) himself was aware of the impossibility of reswitching due to the full determinacy (or over-determinacy) that arises when the number of inputs is equal to (or lower than) the number of alternative techniques. As Stiglitz (1973, 1974) recognized, this implies that the impossibility of reswitching occurs when the alternative techniques are infinite as in the neoclassical case of a continuous spectrum of input-output coefficients.

All subsequent authors who used this 2-sector, 2-input (1 labor + 1 capital good), 2-technique prototype model where sector 1 produces the consumption good and does not supply inputs to the sector 2 (Samuelson, 1962, p. 205, Hicks, 1965, pp. 139-59; Bruno *et al*., 1966, p. 536; Garegnani, 1970, p. 408; Spaventa, 1970, 1973; Harris, 1973; Sato, 1974; Brown, 1974, 1980; Zarembka, 1975; Gram, 1976) overcame the limitation of no space for reswitching due to the "full determinacy" by introducing assumptions similar to Sraffa's (1960, p. 90): They tried to create the needed under-determinacy by assuming that each technique is associated with a particular typology of the capital good (as claimed earlier, for example, by Taussig, 1939, vol. II, p. 213 and Robinson, 1969a [1956], p. 118) being part of a "book of blueprints." In this interpretation, the configuration is that of a 2-sector, 3-input, 2-technique model where the capital good has different dimensionality and the associated price is not comparable across the two techniques (see Brown, 1980, pp. 380, 414-15, fn. 4 for further discussion). However, the mathematics and numerical solutions coincide with that of a 2-2-2 model, although the interpretation of the results is different. Indeed, nobody among the authors cited above endeavored numerical examples of reswitching of such a model except Garegnani (1970, p. 408), whose peculiarities are shown in the case study no. 1 reported below.

*Case study no. 1*: Illustration of case A of the equal number of inputs and techniques using Garegnani's (1970) numerical example. —A selection of seven numeric examples was presented by Garegnani (1970) in a series of bilateral comparisons in the framework of the 2-2-2 Sraffian model. The aim was to show "how far the relation of the rate of interest and the value of capital per worker in the production of a commodity can differ from what traditional theory postulates" (*ibid*., p. 428). The input-output coefficients of labor and capital goods are defined as continuous functions of a parameter u except the constant labor coefficient (set equal to 1) of the sector producing the capital good. The labor coefficient in the production of consumption goods and the capital coefficient in the production of capital goods are increasing functions of *u*, whereas the capital coefficient in the production of consumption goods is decreasing in *u* as shown in **Table 1**. Taking seven values of *u* in increasing order from 0 to 1.505, Garegnani compared the simultaneous solutions of the model with corresponding seven alternative techniques. The external envelope of the real wage-profit rate curves would suggest the reswitching of techniques along the resulting frontier.

**TABLE 1**—GAREGNANI'S (1970, P. 429) FAMILY OF COEFFICIENTS DEFINED PARAMETRICALLY

| Technique | Parameter *u* | $a_{L1}$ | $a_{K1}$ | $a_{L2}$ | $a_{K2}$ |
|---|---|---|---|---|---|
| *I* | 0.000 | 0.500 | 0.750 | 1 | 0.833 |
| *II* | 0.250 | 2.504 | 0.424 | 1 | 0.839 |
| *III* | 0.500 | 3.930 | 0.237 | 1 | 0.845 |
| *IV* | 0.750 | 4.834 | 0.133 | 1 | 0.851 |
| *V* | 1.000 | 5.478 | 0.075 | 1 | 0.857 |
| *VI* | 1.250 | 5.974 | 0.042 | 1 | 0.863 |
| *VII* | 1.505 | 6.391 | 0.023 | 1 | 0.868 |

Garegnani stated:

> [T]he cheaper system will be the same at both wage rates and price systems. Moreover, the tendency of producers to switch to whichever system is cheaper in the existing price situation will bring them to the system giving the highest $w_L$; while systems giving the same $w_L$ for the same *r* will be indifferent and can co-exist (*Ibid*., p. 411).

This contention can be contrasted with the resulting relative levels of total costs of production. In order to save space, we now take only the intermediate cases of $u$ = 0.75 and $u$ = 0.50, respectively, with the depreciation rate $\delta = 1$ of the capital good in both techniques. The numerical solutions are shown in **Table 2**.

Technique *III* (u = 0.50)

$$\mathbf{A}_W^{III} = \begin{array}{c|cc} & \text{Consumption good} & \text{Capital good} \\ \text{Labour input} & 3.930 & 1.000 \\ \text{Capital good input} & 0.237 & 0.845 \end{array}$$

Technique *IV* (u = 0.75)

$$\mathbf{A}_W^{IV} = \begin{array}{c|cc} & \text{Consumption good} & \text{Capital good} \\ \text{Labour input} & 4.834 & 1.000 \\ \text{Capital good input} & 0.133 & 0.851 \end{array}$$

with the depreciation rate $\delta = 1$ of the capital good in both techniques. The numerical solutions are shown in **Table 2**. This table implies homogeneous labor and capital good across the techniques. This assumption makes it possible to construct comparable Laspeyres- and Paasche-type cost indexes across the techniques under different relative price conditions. The two techniques appear to be equally "profitable" when yielding the same real rental of the capital goods for a given real wage although leading to different interest rates and different real purchase prices of capital goods. Similarly, the two techniques may yield the same interest rate but a different degree of profitability measured by the real rental of the capital good. Such a case confutes the general validity of the Sraffian identification of the most "profitable" technique as the one which affords the highest interest rate for a given real wage.

Under the first interpretation that the capital good has the same typology in both techniques, **Table 2** contains three solutions. The middle line marked with double stars refers to a stationary equilibrium with common relative factor prices but different interest rates with the two techniques. The system of equations (26) becomes:

$$\frac{1}{w_L^{**}} w_K^{**} (a_{K1}^{IV} - a_{K1}^{III}) = -(a_{L1}^{IV} - a_{L1}^{III})$$

TABLE 2—SOLUTIONS WITH TWO TECHNIQUES *III* AND *IV* USING GAREGNANI'S (1970, P. 429) NUMERICAL EXAMPLE

| $\frac{w_L}{p_1}$ | $r$ |  | $w_K / p_1$ |  | $w_K / w_L$ |  | Technique in use | $\frac{C^{(NU)}}{C^{(U)}}$ (Paasche) | $\frac{C^{(NU)}}{C^{(U)}}$ (Laspeyres) |
|---|---|---|---|---|---|---|---|---|---|
|  | *III* | *IV* | *III* | *IV* | *III* | *IV* |  |  |  |
| 0.041* | 0.169 | 0.169 | 3.544 | 6.045 | 86.439 | 147.447 | *IV* | 1.495 | 1.590 |
| 0.1669** | 0.0416 | 0.0351 | 1.451 | 1.451 | 8.692 | 8.692 | *III - IV* | 1.000 | 1.000 |
| 0.169* | 0.041 | 0.041 | 1.422 | 1.219 | 8.414 | 7.215 | *III* | 1.005 | 1.027 |

*NU*: Technique not in use*; U*: technique in use.
* Switch point indicated by Garegnani (1970, p. 429).
** Actual switch point with equal relative rental prices, but different levels of *r* and different real prices of capital goods with two alternative techniques.
*Note*: The Paasche and Laspeyres cost indexes $C^{(NU)} / C^{(U)}$ are respectively weighted with the input-output coefficients of the technique "not in use" and the technique "in use."
This table implies homogeneous labor and capital good across the techniques. This assumption makes it possible to construct comparable Laspeyres and Paasche cost indexes across the techniques under different relative price conditions. The two techniques appear to be equally "profitable" when yielding the same real rental of the capital goods for a given real wage although leading to different interest rates and different real purchase prices of capital goods. Similarly, the two techniques may yield the same interest rate but different degree of profitability measured by the real rental of the capital good. Such a case confutes the general validity of the Sraffian identification of the most "profitable" technique with the one which affords the highest interest rate for a given real wage.

whereby $\frac{1}{w_L^{**}} w_K^{**}$ is the rental price relative to wage. Only one switch point occurs although with different levels of the interest rate across the techniques at the given real wage.

The first and the third line marked with a single star in Table 2 refer to two different situations with the same real wage and interest rates but different relative factor prices. No genuine Sraffian reswitching point occurs where the real wage, interest rate, and relative factor prices are simultaneously the same with the two techniques. This example shows that a technique may be the most "profitable" for a given level of real wage in terms of sustainability of a capital good cost but not necessarily in terms of the capacity of paying the highest interest rate.

Under the second interpretation, which was adopted by Garegnani (1970), each technology uses and produces a different typology of capital good. The genuine Sraffian reswitching occurs in the coordinate space of real wage and interest rate. In a binary comparison of techniques, the system of equations (26) becomes:

$$\frac{1}{w_L^*}\mathbf{w}_K^*(\mathbf{a}_{Ki}^{IV} - \mathbf{a}_{Ki}^{III}) = -(a_{Li}^{IV} - a_{Li}^{III}) \qquad \text{for sector } i = 1,2$$

that is

$$\frac{1}{w_L^*}\mathbf{w}_{Ki}^*(\mathbf{a}_{Ki}^{IV} - \mathbf{0}) + \frac{1}{w_L^*}\mathbf{w}_{Ki}^*(\mathbf{0} - \mathbf{a}_{Ki}^{III}) = -(a_{Li}^{IV} - a_{Li}^{III}) \qquad \text{for sector } i = 1,2$$

where $\frac{1}{w_L^*}w_{Ki}^*$ and $\frac{1}{w_L^*}w_{Ki}^*$ refer, in this interpretation, to two different types of capital goods. These equations are indeed fully consistent with the pure marginalist theory of relative factor rewards.

### B. *The case of the number of techniques being less than to the number of factor inputs*

Confining the discussion to the choice between pairs of techniques and noting that the determinant of the Sraffian inverse matrix of the system (9) implies a polynomial whose possible maximum degree is equal to the number of sectors, the following theorem has been established:

THEOREM: Maximum number of genuine switch points (Bruno et al., 1966, p. 542): (1) *In the general* n-*sector model there may be up to* n *switching points between any two techniques, and thus a technique may recur up to (*n - *1) times. (2) "Adjacent" techniques on two sides of a switching point will usually differ from each other only with respect to one activity. Techniques, in general, may differ with respect to* m *activities (*n ≥ m *> 1) only if certain m independent* n-*th degree polynomials happen to have a common root at that switching point.*

This theorem implies that, in the case of two sectors, three inputs, and two different techniques, the maximum number of genuine switch points in the coordinate space of real wage and interest rate is equal to 2. These switch points can be mapped into the respective corresponding locus of linear intersection of the two technique planes in the 3-dimensional coordinate space of the real factor prices. Let the system of equations be represented as

$$a_{L1}^{I}\frac{w_L}{p_1}+a_{K11}^{I}\frac{w_{K1}}{p_1}+a_{K21}^{I}\frac{w_{K2}}{p_1}=1$$
$$a_{L1}^{II}\frac{w_L}{p_1}+a_{K11}^{II}\frac{w_{K1}}{p_1}+a_{K21}^{II}\frac{w_{K2}}{p_1}=1 \tag{30}$$

As two (dependent) real input-prices are functions of a third (independent) real input price and, given that $\frac{w_{K1}}{p_1}=\frac{p_1(1+r)}{p_1}=1+r$, the previous equation system has the following set of solutions written in matrix form:

$$\begin{bmatrix}\frac{w_L}{p_1} & \frac{w_{K2}}{p_1}\end{bmatrix}\cdot\begin{bmatrix}a_{L1}^{I} & a_{L1}^{II}\\ a_{K21}^{I} & a_{K21}^{II}\end{bmatrix}+(1+r)\begin{bmatrix}a_{K11}^{I} & a_{K11}^{II}\end{bmatrix}=\begin{bmatrix}1 & 1\end{bmatrix} \tag{31}$$

which is the equation of a straight line in a three-dimensional space as that passing through the points A and B in **Figure 3**.

Let us consider two sub-cases regarding the matrix singularity.

*Case $B1$: The matrix* $\begin{bmatrix}a_{L1}^{I} & a_{L1}^{II}\\ a_{K21}^{I} & a_{K21}^{II}\end{bmatrix}$ *in (31) is singular*. —

The matrix singularity implies that the real factor-price hyperplane of the focused sector is the same with the two techniques. The common factor-price equation is derived:

$$\frac{w_{K2}}{p_1}=1-\frac{a_{L1}}{a_{K21}}\cdot\frac{w_L}{p_1}+\frac{a_{K11}}{a_{K21}}(1+r) \tag{32}$$

with

$$a_{L1}=a_{L1}^{I}=a_{L1}^{II}$$
$$a_{K11}=a_{K11}^{I}=a_{K11}^{II}$$
$$a_{K21}=a_{K21}^{I}=a_{K21}^{II}$$

In this special case, all common levels of the interest rate and real wage rate are sufficient for the two techniques to determine the same real factor price level thus satisfying Sraffa's requirements for the identification of switch points. Almost all the numerical examples proposed by Sraffa's followers were formulated within this special case. The reduced form of the foregoing equation is obtained for the relative price level.

$$\frac{p_2^{(T)}}{p_1^{(T)}} = \frac{a_{L2}^{(T)} + (a_{L1}a_{K12}^{(T)} - a_{L2}^{(T)}a_{K11})(1+r)}{a_{L1} + (a_{L2}^{(T)}a_{K21} - a_{L1}a_{K22}^{(T)})(1+r)} \quad \text{for } T = I, II \tag{33}$$

and for the real wage assumed to be paid *post factum*

$$\frac{w_L^{(T)}}{p_1^{(T)}} = \frac{1-(a_{K11}+a_{K22}^{(T)})(1+r)+(a_{K11}a_{K22}^{(T)} - a_{K21}a_{K12}^{(T)})(1+r)^2}{a_{L1} + (a_{L2}^{(T)}a_{K11} - a_{L1}a_{K22}^{(T)})(1+r)} \quad \text{for } T = I, II \tag{34}$$

which can be converted to the real wage paid *ante factum* $w_0^{(T)} / p_1$ used in many numerical examples cited below according to the formula $\frac{w_0^{(T)}}{p_1} = \frac{w_L^{(T)}}{p_1}(1+r)^{-1}$.

Genuine reswitching is then possible in the real wage-interest space with common levels $r^*$ and equalities of relative prices $\frac{w_L^{(I)}}{p_1} = \frac{w_L^{(II)}}{p_1}$ and $\frac{p_2^{(I)}}{p_1} = \frac{p_2^{(II)}}{p_1}$. Reswitching appears as a response to the return to previous ranking levels of the input price ratios

$$\begin{aligned}
\frac{w_L^{(II/I)}}{p_1^{(II/I)}} &\equiv \frac{w_L^{(II)}}{p_1^{(II)}} / \frac{w_L^{(I)}}{p_1^{(I)}} \\
\frac{w_{Ki}^{(II/I)}}{p_1^{(II/I)}} &\equiv \frac{w_{Ki}^{(II)}}{w_L^{(II)}} / \frac{w_{Ki}^{(I)}}{w_L^{(I)}} \qquad \text{for } i = 1, 2 \\
\frac{w_{Ki}^{(II/I)}}{w_L^{(II/I)}} &\equiv \frac{w_{Ki}^{(II)}}{w_L^{(II)}} / \frac{w_{Ki}^{(I)}}{w_L^{(I)}} \qquad \text{for } i = 1, 2
\end{aligned} \tag{35}$$

Given (10), (11), and (12), these ratios are a function of *r* for the given compared techniques $\mathbf{A}_T^{(I)}$ and $\mathbf{A}_T^{(II)}$.

*Case B2: The matrix* $\begin{bmatrix} a_{L1}^I & a_{L1}^{II} \\ a_{K21}^I & a_{K21}^{II} \end{bmatrix}$ *is not singular*. —

Solving the system (26) yields

$$\begin{bmatrix} \frac{w_L}{p_1} & \frac{w_{K2}}{p_1} \end{bmatrix} = \begin{bmatrix} 1 & 1 \end{bmatrix} \begin{bmatrix} a_{L1}^{I} & a_{L1}^{II} \\ a_{K21}^{I} & a_{K21}^{II} \end{bmatrix}^{-1} - (1+r) \begin{bmatrix} a_{K11}^{I} & a_{K11}^{II} \end{bmatrix} \begin{bmatrix} a_{L1}^{I} & a_{L1}^{II} \\ a_{K21}^{I} & a_{K21}^{II} \end{bmatrix}^{-1} \tag{36}$$

This defines the linear locus of points of two intersecting planes in the 3-dimensional coordinate space of real factor prices. This locus of points is the set of solution values corresponding to given levels of the profit rate. Since the two planes are linear, they share *only one* intersecting straight line such as that shown by the AB line in **Figure 3**. This implies that common levels of the profit rate and real wage are not sufficient for determining the switch points as these should occur only under the *very special condition* of mapping into the intersecting straight line within the coordinate space of relative input prices.

*Case studies nos. 2 to 7: Numerical examples of the two-sector model with three inputs and two techniques. —*

Various numerical examples of case *B*1 have been proposed in the literature, including those shown in **Table 3A** containing a synoptic collection of the coefficients of the equations (33) and (34) used in five well-known contributions, where the techniques differ only in the capital good producing sector.

A numerical example of case *B*2 has been provided by Pertz (1980), considered in the case study no. 7, where both techniques differ in both sectors using and producing different types of capital goods as shown in **Table 3B**. Similarly to Garegnani's (1970, p. 429 ) numerical example of Case *A*, the two matrices contain data that pertain to different factor inputs.

The technique *I* has a lower capital per unit of labor than technique *II* in all the numerical examples considered here, except Bruno *et al.* (1966) where the relative capital intensity is reversed between the two techniques. Given the level of profit rate, the two techniques are compared in terms of real wages and relative prices of commodity inputs in **Table 4**.

All cited authors report the comparisons of the two techniques based only on the real wage values for given levels of the profit rate (reported, respectively, in the second and first columns) without checking the full range of the relative prices (reported in the third column). As noted, the cost-minimizing technical choice for a given profit rate is affected by relative input prices. In all the numerical instances of reswitching, the choice of the more (less) capital-intensive technique is invariably associated with the lower (higher) capital rental price in terms of labor. Therefore, all the well-known counterexamples re-examined here appear entirely consistent with the expectations of the marginalist theory of cost-minimizing choices of techniques.

Given such results, the Sraffians' critical interpretation of the switch points is disproved on the ground of the non-monotonic effects of relative input prices resulting from monotonic changes in the rate of profit. All the Sraffian analysts failed to recognize that reswitching over the range of interest rate happens when facing *non-monotonic* changes (*decreasing* and then *increasing* or vice versa) in relative rental prices of physical capital induced by *monotonic* changes in the profit rate. As an illustration, **Figure 4** shows the ratios $w_{Ki}^{(II/I)} / w_{L}^{(II/I)}$ defined by (35) for the rental prices of both capital goods relative to the wage in correspondence of a range of given levels of the profit rate using the coefficients of Garegnani's (1976, pp. 425-426) numerical example. The results graphically demonstrate how the differential decrease and subsequent increase in relative capital price in terms of labor over monotonic increases in the profit rate would bring the cost-minimizion from technique *I* to technique *II* and then turning back to technique *I*. A similar pattern could be observed in all the other numerical counterexamples shown in **Table 4**.

**TABLE 3A**—INPUT-OUTPUT COEFFICIENTS IN NUMERICAL EXAMPLES OF CASE STUDIES NOS. 2 TO 6

| | Case no. 2: Bruno *et al.* (1966, p. 537) | Case no 3: Garegnani (1966, p. 566, fn.1) | Case no. 4: Garegnani (1976, pp. 425-426) | Case no. 5: Sato (1976, p. 428) | Case no. 6: Laibman&Nell (1977, p. 881) |
|---|---|---|---|---|---|
| $\delta$ | 1.00 | 1.00 | 1.00 | 1.00 | 1.00 |
| Common technique in the focused sector 1 | | | | | |
| $a_{L1}$ | 1.00 | 8.90 | 1.00 | 1.00 | 1.00 |
| $a_{K11}$ | 0.00 | 0.00 | 1/12 | 0.20 | 0.10 |
| $a_{K21}$ | 0.10 | 379/423 | 1/3 | 0.40 | 1.00 |
| Alternative techniques in sector 2 | | | | | |
| *Technique I* | | | | | |
| $a_{L2}$ | 0.66 | 9/50 | 1.0 | 1.5 | 0.558720 |
| $a_{K21}$ | 0.02 | 1/2 | 1/6 | 0.4 | 0.135872 |
| $a_{K22}$ | 0.30 | 0.1 | 1/6 | 0.2 | 0.358720 |
| *Technique II* | | | | | |
| $a_{L2}$ | 0.01 | 3/2 | 92/91 | 1.55 | 0.567120 |
| $a_{K21}$ | 0.71 | 1/4 | 137/546 | 0.5205 | 0.261712 |
| $a_{K22}$ | 0.00 | 5/12 | 19/273 | 0.08 | 0.117120 |

(*) The notation of input-output coefficients indicated for Bruno *et al*. (1966, p. 537) is reversed here to make it consistent with the other cases where the focused sector is the first one.

**TABLE 3B**—INPUT-OUTPUT COEFFICIENTS IN THE NUMERICAL EXAMPLE OF CASE STUDY NO. 7: PERTZ (1980, P. 1016)

Technique *I*, sector 1 2:

$$\mathbf{a}_L^I = [0.8 \quad 1.0]$$
$$\mathbf{a}_{K1}^I = [0.0 \quad 0.1]$$
$$\mathbf{a}_{K2}^I = [0.7 \quad 0.0]$$
$$\delta = 1.0$$

Technique *II*, sector 1 2:

$$\mathbf{a}_L^{II} = [1.3 \quad 0.9]$$
$$\mathbf{a}_{K1}^{II} = [0.0 \quad 0.145]$$
$$\mathbf{a}_{K2}^{II} = [0.5 \quad 0.0]$$
$$\delta = 1.0$$

The technique *I* has a lower capital per labor unit than technique *II* in all the numerical examples considered here, except Bruno *et al.* (1966) where the relative capital intensity is reversed between the two techniques. Given the level of profit rate, the two techniques are compared in terms of real wages and relative prices of commodity inputs in **Table 4**.

All cited authors report the comparisons of the two techniques based only on the real wage values for given levels of the profit rate (reported, respectively, in the second and first columns) without checking the full range of the relative prices. As noted, the cost-minimizing technical choice for a given profit rate is affected by relative input prices. In all the numerical instances of reswitching, the choice of the more (less) capital-intensive technique is invariably associated with the lower (higher) capital rental price in terms of labor. Therefore, all the well-known counterexamples re-examined here appear entirely consistent with the expectations of the neoclassical theory of cost-minimizing choices of techniques.

Given such results, the Sraffians' critical interpretation of the switch points is disproved on the ground of the non-monotonic effects of relative input prices resulting from monotonic changes in the rate of profit. All the Sraffian analysts failed to recognize that reswitching over the range of interest rate happens when facing *non-monotonic* changes (*decreasing* and then *increasing* or vice versa) in relative rental prices of physical capital induced by *monotonic* changes in the profit rate. As an illustration, **Figure 4** shows the ratios $w_{Ki}^{(II/I)} / w_{L}^{(II/I)}$ defined by (35) for the rental prices of both capital goods relative to the wage in correspondence of a range of given levels of the profit rate using the coefficients of Garegnani's (1976, pp. 425-426) numerical example. The results graphically demonstrate how the differential decrease and subsequent increase in relative capital price in terms of labor over monotonic increases in the profit rate would bring the cost-minimizion from technique *I* to technique *II* and then turning back to technique *I*. A similar pattern could be observed in all the other numerical counterexamples shown in **Table 4**.

**TABLE 4**—"RESWITCHING" AS A RESPONSE TO THE EFFECT OF INCOME DISTRIBUTION ON RELATIVE PRICES

| $r$ | $\frac{w_L^{(II/I)}}{p_1^{(II/I)}}$ | $\frac{w_{K2}^{(II/I)}}{p_1^{(II/I)}}$ | $\frac{w_{K1}^{(II/I)}}{w_L^{(II/I)}}$ | $\frac{w_{K2}^{(II/I)}}{w_L^{(II/I)}}$ | Technique in use |
|---|---|---|---|---|---|
| (Results shown by the authors) | | (Results not shown by the authors) | | | |
| Bruno *et al.* (1966, p. 537) | | | | | |
| 0.250000 | 1.010214 | 0.925823 | 0.989890 | 0.916462 | *II* |
| 0.465809* | 1.000000 * | 1.000000* | 1.000000 * | 1.000000* | *I-II* |
| 1.000000 | 0.969773 | 1.084631 | 1.031168 | 1.118436 | *I* |
| 1.668760* | 1.000000 * | 1.000000* | 1.000000 * | 1.000000* | *I-II* |
| 1.857000 | 1.097942 | 0.939496 | 0.910794 | 0.855688 | *II* |
| Garegnani (1966, p. 566) | | | | | |
| 0.010000 | 0.994627 | 1.005000 | 1.005402 | 1.010430 | *I* |
| 0.100000* | 1.000000 * | 1.000000* | 1.000000 * | 1.000000* | *I-II* |
| 0.150068 | 1.001467 | 0.999320 | 0.998575 | 0.997896 | *II* |
| 0.200000* | 1.000000 * | 1.000000* | 1.000000 * | 1.000000* | *I-II* |
| 0.250643 | 0.991743 | 1.001989 | 1.008326 | 1.010462 | *I* |
| Garegnani (1976, pp. 425-26) | | | | | |
| 0.100000 | 0.999785 | 1.000363 | 1.000215 | 1.000579 | *I* |
| 0.333333* | 1.000000 * | 1.000000* | 1.000000 * | 1.000000* | *I-II* |
| 0.450000 | 1.000020 | 0.999979 | 0.999980 | 0.999959 | *II* |
| 0.500000* | 1.000000 * | 1.000000* | 1.000000 * | 1.000000* | *I-II* |
| 0.900000 | 0.998604 | 1.000713 | 1.001398 | 1.002112 | *I* |
| Sato (1976, pp. 428-30) | | | | | |
| 0.000000 | 0.999621 | 1.000284 | 1.000379 | 1.000664 | *I* |
| 0.038000* | 1.000000 * | 1.000000* | 1.000000* | 1.000000* | *I-II* |
| 0.250000 | 1.002088 | 0.999197 | 0.997916 | 0.997114 | *II* |
| 0.595000* | 1.000000* | 1.000000* | 1.000000* | 1.000000* | *I-II* |
| 0.639000 | 0.986951 | 1.000263 | 1.013222 | 1.013489 | *I* |
| Laibman and Nell (1977, p. 881) | | | | | |
| 0.100000 | 0.999703 | 1.000156 | 1.000298 | 1.000454 | *I* |
| 0.200000* | 1.000000 * | 1.000000* | 1.000000* | 1.000000* | *I-II* |
| 0.300000 | 1.000175 | 0.999952 | 0.999826 | 0.999777 | *II* |
| 0.400000* | 1.000000 * | 1.000000* | 1.000000* | 1.000000* | *I-II* |
| 0.500000 | 0.998454 | 1.000132 | 1.001548 | 1.001680 | *I* |
| Pertz (1980, p. 1016) | | | | | |
| 0.500000 | 0.930455 | 1.074743 | 1.074742 | 1.155072 | *I* |
| 1.170000* | 1.000000 * | 1.000000* | 1.000000* | 1.000000* | *I-II* |
| 1.750000 | 1.030747 | 0.970167 | 0.970170 | 0.941227 | *II* |
| 2.250000* | 1.000000* | 1.000000* | 1.000000* | 1.000000* | *I-II* |
| 2.500000 | 0.887452 | 1.126773 | 1.126821 | 1.269672 | *I* |

* Sraffian switch point; *Note*: The ratio of the real rental price $w_{K1}^{(II/I)} / p_1^{(II/I)} = (1+r)/(1+r) = 1$ (with the same levels of $r$ being conjecturally predetermined for both techniques) is omitted in this table.

Given such results, the Sraffians' critical interpretation of the switch points is disproved on the ground of the non-monotonic effects of relative input prices resulting from monotonic changes in the rate of profit. All the Sraffian analysts, for example, Pasinetti's (1966, p. 514), failed to recognize that reswitching over the range of interest rate happens when facing *non-monotonic* changes (*decreasing* and then *increasing* or vice versa) in relative rental prices of physical capital induced by *monotonic* changes in the profit rate. As an illustration, **Figure 4** shows the ratios $w_{Ki}^{(II/I)} / w_{L}^{(II/I)}$ defined by (35) for the rental prices of both capital goods relative to the wage in correspondence of a range of given levels of the profit rate using the coefficients of Garegnani's (1976, pp. 425-426) numerical example. The results graphically demonstrate how the differential decrease and subsequent increase in relative capital price in terms of labor over monotonic increases in the profit rate would bring the cost-minimizing choices of techniques from a point A beyond a point B while passing from technique *I* to technique *II* and then going back to technique *I*. A similar pattern could be observed in all the other numerical counterexamples shown in **Table 4**.

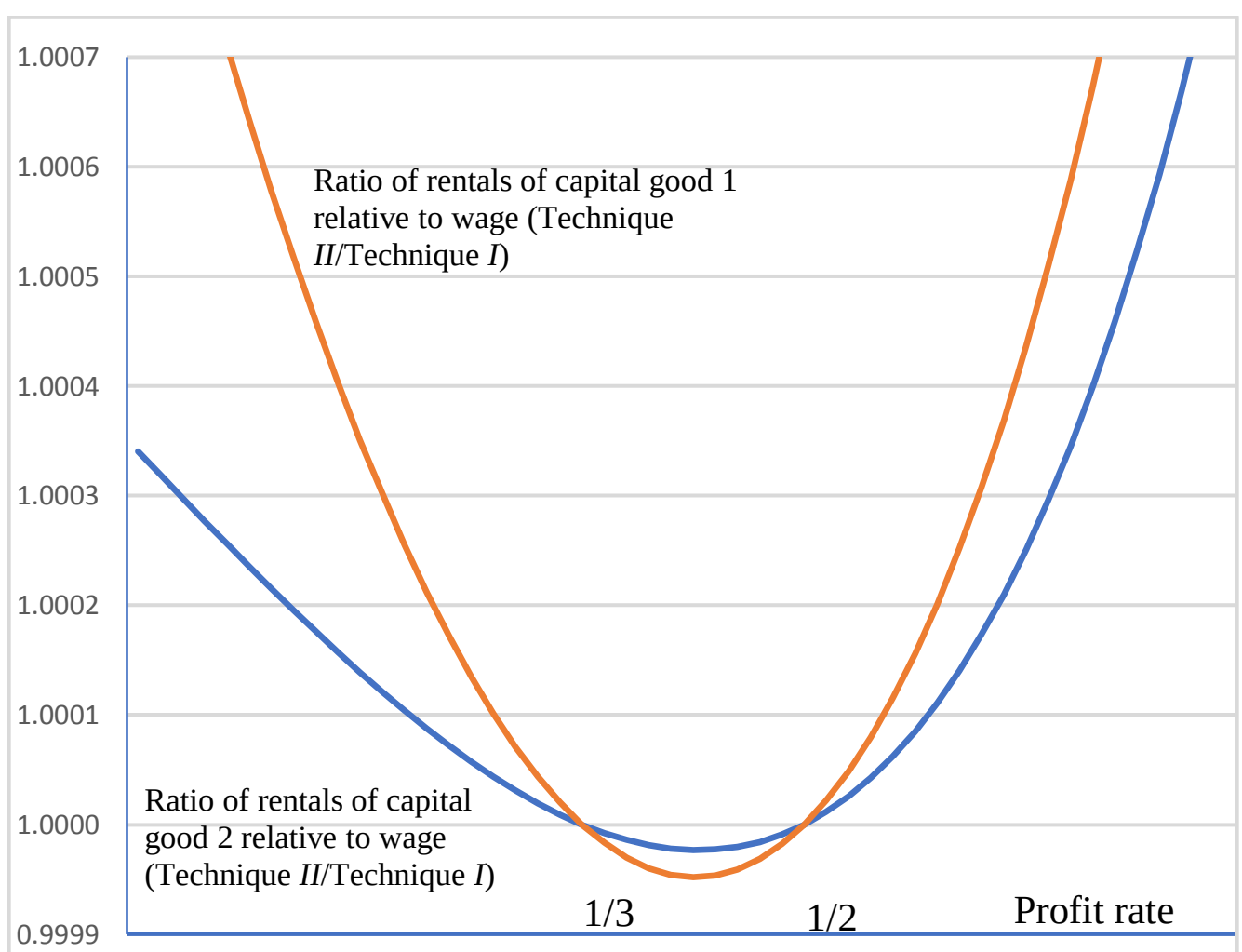

**FIGURE 4**—NON-LINEAR EFFECTS OF THE PROFIT RATE ON RELATIVE PRICES (USING GAREGNANI'S 1976, PP. 425-426 NUMERICAL EXAMPLE)

If all these changes in relative factor prices are sorted in increasing order as in **Table 5,** then the associated order of the "technique in use" reveals no paradoxical reswitching. In Garegnani's (1976) general case, the levels of relative prices in correspondence of two roots for the interest rate of a polynomial equation. With these multiple levels of interest rate in combination with the given wage rates, the two techniques may or may not coexist with the same relative prices. Most importantly, as long as the rental prices of capital goods relative to wage are lower in the capital-intensive technique *II* than in labor-intensive technique *I*, the former remains in use. The opposite happens when rental-wage ratios in technique *II* are higher than in technique *I*. Therefore, in all Sraffian models, monotonic changes in relative factor prices give rise to monotonic changes of techniques in full consistency with the pure marginalist theory of factor rewards. Similar results invariably occur in all the other cases mentioned in **Table 4**.

The cost comparisons of cases with heterogeneous capital goods, such as those considered by Pertz's (1980, p. 1016) are problematic as the relative factor prices across different techniques cannot be immediately constructed. Solutions could be found as in the general cases of index numbers in the presence of new products. In any case, these solutions are consistent with the system of marginal rate equations (26) similarly to those obtained above with Garegnani's (1970) example. However, **Table 4** shows the relative factor prices across the two techniques also for Pertz's example on the assumption that the capital goods have the same typology across the techniques similar to those of the other examples. The wage-interest solutions are the same as those obtained by Pertz himself while those regarding the relative factor prices indicate the expected negative relation between relative factor-intensities and relative factor prices.

Indeed, Sraffa (1960) insisted on the fact that, in his intersectoral model, changes in the profit or interest rate may affect the relative prices significantly and non-linearly. His analysis confirmed the old discovery that the distribution of income generally yields non-linear effects on relative prices and the internal structure of production. However, the U-turn changes of such effects on relative prices and their implications for the interpretation of technical choices were overlooked and remained hidden in the reswitching debate. In light of the present solution, the "reswitching paradox" cannot be deemed anymore as a contradiction of the neoclassical paradigm.

**TABLE 5**—TECHNIQUES IN USE OVER THE RANGE OF RELATIVE FACTOR PRICES (USING DATA FROM GAREGNANI'S 1976, PP. 425-426)

| Relative rental price 1 $\frac{w_{K1}^{(II/I)}}{w_L^{(II/I)}}$ | Relative rental price 2 $\frac{w_{K2}^{(II/I)}}{w_L^{(II/I)}}$ | Rate of interest $r$ | Technique in use | Cost ratio $\frac{C^{(NU)}}{C^{(U)}}$ (per cent) |
|---|---|---|---|---|
| 0.99998* | 0.99996* | 0.40* | *II* | 100.001* |
| **1.00000**** | **1.00000**** | **1/3 and 1/2**** | ***I* - *II***** | **100.00**** |
| 1.00022 | 1.00058 | 0.1 and 0.65 | ***I*** | 100.02 |
| 1.00140 | 1.00211 | (negative) and 0.90 | *I* | 100.14 |

* Maximum cost advantage of technique *II*.
**Unique switch point in the coordinate space of real factor prices
*Note*: The cost ratio $C^{(NU)}/C^{(U)}$ is computed on costs simulated with the relative factor prices obtained, respectively with the technique "not in use" (*NU*) and the technique "in use" (*U*).

## 7. Reverse capital deepening

The reversing of the relative intensity of *financial* capital or *capital value* in the overall economy under stationary equilibrium with monotonic changes of the interest rate was already noted by Fisher (1907) (see Samuelson, 1966, and Velupillai, 1975, 1995 on Fisher's "discovery") and independently by Robinson (1953) and Champernowne (1953) (see Harcourt, 1972, pp. 124-76, Burmeister, 2008, Kurz, 2008, and Scazzieri, 2008 for further discussions).

The relative producer cost, say in terms of the *j*th commodity, of capital goods per unit of labor in our model is given by (assuming for simplicity a null depreciation rate):

$$r \cdot \frac{p_{Ki}}{p_j} \cdot \frac{a_{Kij}}{a_{Lj}} = \frac{w_{Ki}}{p_j} \cdot \frac{a_{Kij}}{a_{Lj}} \tag{37}$$

In the right-hand side of the foregoing equality, the “price” component *r* (the rate of interest or rate of profit) multiplies the “deflated” component $\frac{p_{Ki}}{p_j} \cdot \frac{a_{Kij}}{a_{Lj}}$, which can be interpreted as the value of the *i*th capital goods per unit of labor in terms of the *j*th commodity. This value is obtained by evaluating the physical capital goods per unit of labor $a_{Ki}$ with the relative price $p_{Ki} / p_j$. The cost of the *i*th capital good is decomposed in the right-hand side of (37) in terms of the relative rental price of capital goods $w_{Ki} / p_j$ and the ratio of capital to labor technical coefficients $a_{Kij} / a_{Lj}$. In the intersectoral model, the effects of *r* on these two variables are called respectively “price” and “real” Wicksell effects. The reverse deepening in capital value can occur when the “price” Wicksell effect overcome the opposite real Wicksell effect.

## 8. Conclusion

The reswitching of techniques in the Sraffian intersectoral model of a cost-minimizing economy over a range of interest or real wage turns out to be misinterpreted as a paradox in capital theory. The Sraffian analysis of an intersectoral production economy has stressed two important points: 1) monotonic changes in the interest rate (or real wage) affect relative prices non-linearly; 2) in a genuine switch point, alternative techniques share the same system of relative prices. Drawing on these results, the present article has solved the paradox by showing that the reswitching of techniques can be rationalized as a response to a U-shaped turn in the factor prices ratios across the technques over different levels on interest or wage rate.

The misconception of “reswitching paradox” has remained unchallenged as long as the existence of multiple roots for interest rate of the equality of price ratio across the techniques—already well-known in financial engineering—has been disregarded in the Sraffian analysis. Also known, well ahead of the Cambridge controversy, was that the essence of capital theory resides in the relationship between the interest rate and the price structure. By taking into account of this knowledge, the cost-minimizing changes of techniques in the Sraffian model always turn out to be consistent with the marginalist expectations of the negative correlation of input prices and quantities even in cases of reswitching in the wage-interest space.

## APPENDIX

**Proposition A1**. **Equivalent formulations of intertemporal minimization of costs:** *Let us assume the absence of technical progress, no internal adjustment costs, and static expectations regarding the output quantity and prices of output and inputs. The following two alternative formulations of the minimization of the present value of total costs of production in stationary equilibrium over the entire intertemporal program are equivalent:*

$$\begin{aligned}\mathrm{Min}_{k,\dot{k}} \Big\{ & \textstyle\sum_{t=0}^{t=\infty} (1+r)^{-t} \left[ C_v(\mathbf{w}_v; k_t, y, t) + p_k \delta k_t \right] \\ & + \textstyle\sum_{t=0}^{t=\infty} (1+r)^{-t} \int_{\tau=0}^{\tau=1} e^{-\rho\tau} p_k \dot{k}_{t+\tau} \cdot d\tau \Big\}\end{aligned} \tag{A1}$$

$$\begin{aligned}\mathrm{Min}_{k} & \Big\{ \textstyle\sum_{t=0}^{t=\infty} (1+r)^{-t} \left[ C_v(\mathbf{w}_v; k_t, y, t) + p_k (\delta + r) k_t \right] \Big\} \\ & = \Big[ \textstyle\sum_{t=0}^{t=\infty} (1+r)^{-t} C(\mathbf{w}; y, t) \Big] - p_k k_0\end{aligned} \tag{A2}$$

*where* $C_v(\mathbf{w}_v; k_t, y, t) \equiv \mathrm{Min}_{\mathbf{x}_v} \{ \mathbf{w}_v \mathbf{x}_v : T(\mathbf{x}_v, k_t, y, t) \}$ *is the minimum restricted or variable cost function subject to the given convex technology* $T(\cdot)$ *and output level y at period t, with* $\mathbf{w}_v$ *and* $\mathbf{x}_v$ *being, respectively, the vectors of factor prices and quantities of variable factor services (including labor services), and* $C(\mathbf{w}; y, t) \equiv \mathrm{Min}_{\mathbf{x}} \{ \mathbf{w} \cdot \mathbf{x} : T(\mathbf{x}_v, k_t, y, t) \}$ *is the minimum total cost function when all inputs are variable so that* $\mathbf{x} \equiv [\mathbf{x}_v \quad k_t]$, *and*

$$w_k = p_K (\delta + r) \tag{A3}$$

*is the user cost or rental price of the capital good,* $k_t$ *is the quantity of capital good in period t,* $\dot{k}_{t+\tau}$ *is its net increment in period* $t+\tau$ *,* $\rho = \ln(1+r)$ *where ρ and r are, respectively, the continuous and discrete exponential rates of compounded interest, and* $\delta$ *is the depreciation rate.*

**Proof**: In the last additive term of equation (A1), the integral $\int_{\tau=0}^{\tau=1} e^{-\rho\tau} p_k \dot{k}_{t+\tau} \cdot d\tau$ can be integrated by parts. Setting

$$u_\tau \equiv e^{-\rho\tau}$$

$$v_\tau \equiv p_k \cdot k_{t+\tau}$$

integration by parts is $\int_{\tau=0}^{\tau=1} u \cdot v' = [u \cdot v]_0^1 - \int_{\tau=0}^{\tau=1} u' \cdot v$, and substituting yields

$$\int_{\tau=0}^{\tau=1} e^{-\rho\tau} p_k \dot{k}_{t+\tau} \cdot d\tau = \left[ e^{-\rho} p_k k_{t+1} - p_k k_t \right] - \int_{\tau=0}^{\tau=1} -\rho e^{-\rho\tau} p_k k_{t+\tau} \cdot d\tau$$
$$= \int_{\tau=0}^{\tau=1} e^{-\rho\tau} \rho p_k k_{t+\tau} \cdot d\tau - p_k k_t + e^{-\rho} p_k k_{t+1}$$

Then, the second line of (A1) becomes

$$\sum_{t=0}^{t=\infty} (1+r)^{-t} \int_{\tau=0}^{\tau=1} e^{-\rho\tau} p_k \dot{k}_{t+\tau} \cdot d\tau$$
$$= \sum_{t=0}^{t=\infty} (1+r)^{-t} \int_{\tau=0}^{\tau=1} e^{-\rho\tau} \rho p_k k_{t+\tau} \cdot d\tau + \sum_{t=0}^{t=\infty} (1+r)^{-(t+1)} \cdot p_k k_{t+1} - (1+r)^{-t} \cdot p_k k_t$$
$$= [\sum_{t=0}^{t=\infty} (1+r)^{-t} r p_k k_t] - p_k k_0$$

since

$$e^{-\rho} = (1+r)^{-1};$$

$$\int_{\tau=0}^{\tau=1} e^{-\rho\tau} \rho p_k k_{t+\tau} \cdot d\tau = [-\frac{1}{\rho} \cdot e^{-\rho\tau} \rho p_k k_{t+\tau} \cdot]_0^1 = -(1+r)^{-1} p_k k_{t+1} + p_k k_t$$

Hence, (A1) can be rewritten as (A2). QED.